\documentclass[twocolumn,showkeys,prc]{revtex4}

\usepackage{amsmath}
\usepackage{amssymb}
\usepackage{graphicx}
\usepackage{epsfig}
\usepackage{natbib}

\begin{document}

\title{The relativistic Feynman-Metropolis-Teller theory for white dwarfs in general relativity}

\author{Michael Rotondo}
\affiliation{Dipartimento di Fisica and ICRA, Sapienza Universit\`a di Roma, P.le Aldo Moro 5, I--00185 Rome, Italy}
\affiliation{ICRANet, P.zza della Repubblica 10, I--65122 Pescara, Italy}

\author{Jorge A. Rueda}
\affiliation{Dipartimento di Fisica and ICRA, Sapienza Universit\`a di Roma, P.le Aldo Moro 5, I--00185 Rome, Italy}
\affiliation{ICRANet, P.zza della Repubblica 10, I--65122 Pescara, Italy}

\author{Remo Ruffini}
\email{ruffini@icra.it}
\affiliation{Dipartimento di Fisica and ICRA, Sapienza Universit\`a di Roma, P.le Aldo Moro 5, I--00185 Rome, Italy}
\affiliation{ICRANet, P.zza della Repubblica 10, I--65122 Pescara, Italy}

\author{She-Sheng Xue}
\affiliation{Dipartimento di Fisica and ICRA, Sapienza Universit\`a di Roma, P.le Aldo Moro 5, I--00185 Rome, Italy}
\affiliation{ICRANet, P.zza della Repubblica 10, I--65122 Pescara, Italy}

\date{\today}

\begin{abstract}
The recent formulation of the relativistic Thomas-Fermi model within the Feynman-Metropolis-Teller theory for compressed atoms is applied to the study of general relativistic white dwarf equilibrium configurations. The equation of state, which takes into account the $\beta$-equilibrium, the nuclear and the Coulomb interactions between the nuclei and the surrounding electrons, is obtained as a function of the compression by considering each atom constrained in a Wigner-Seitz cell. The contribution of quantum statistics, weak, nuclear, and electromagnetic interactions is obtained by the determination of the chemical potential of the Wigner-Seitz cell. The further contribution of the general relativistic equilibrium of white dwarf matter is expressed by the simple formula $\sqrt{g_{00}}\mu_{\rm ws}=$ constant, which links the chemical potential of the Wigner-Seitz cell $\mu_{\rm ws}$ with the general relativistic gravitational potential $g_{00}$ at each point of the configuration. The configuration outside each Wigner-Seitz cell is strictly neutral and therefore no global electric field is necessary to warranty the equilibrium of the white dwarf. These equations modify the ones used by Chandrasekhar by taking into due account the Coulomb interaction between the nuclei and the electrons as well as inverse $\beta$-decay. They also generalize the work of Salpeter by considering a unified self-consistent approach to the Coulomb interaction in each Wigner-Seitz cell. The consequences on the numerical value of the Chandrasekhar-Landau mass limit as well as on the mass-radius relation of $^4$He, $^{12}$C, $^{16}$O and $^{56}$Fe white dwarfs are presented. All these effects should be taken into account in processes requiring a precision knowledge of the white dwarf parameters.
\end{abstract}


\maketitle
%
%%%%%%%%%%%%%%%%%%%%%%%%%%%%%%%%%%%%%%%%%%%%%%%%%%%%%%%%%%%%%%
%%%%%%%%%%%%%%%%%%%%%%%%%%%%%%%%%%%%%%%%%%%%%%%%%%%%%%%%%%%%%%
\section{Introduction}\label{sec:1}
%%%%%%%%%%%%%%%%%%%%%%%%%%%%%%%%%%%%%%%%%%%%%%%%%%%%%%%%%%%%%%
%%%%%%%%%%%%%%%%%%%%%%%%%%%%%%%%%%%%%%%%%%%%%%%%%%%%%%%%%%%%%%

The necessity of introducing the Fermi-Dirac statistics in order to overcome some conceptual difficulties in explaining the existence of white dwarfs leading to the concept of degenerate stars was first advanced by R.~H.~Fowler in a classic paper \cite{fowler26}. Following that work, E.~C.~Stoner \cite{stoner29} introduced the effect of special relativity into the Fowler considerations and he discovered the critical mass of white dwarfs \footnote{In doing this, Stoner used what later became known as the exclusion principle, generally attributed in literature to Wolfgang Pauli. For a lucid and scientifically correct historical reconstruction of the contributions to the critical mass concept see \cite{nauenberg08}. For historical details about the exclusion principle see also \cite{heilbron83}.}

\begin{equation}\label{eq:maxmassStoner}
M^{\rm Stoner}_{\rm crit} = \frac{15}{16} \sqrt{5 \pi} \frac{M_{\rm Pl}^3}{\mu^2 m^2_n} \approx 3.72\frac{M_{\rm Pl}^3}{\mu^2 m^2_n} \, ,
\end{equation}
where $M_{\rm Pl} = \sqrt{\hbar c/G} \approx 10^{-5}$ g is the Planck mass, $m_n$ is the neutron mass, and $\mu = A/Z \approx 2$ is the average molecular weight of matter which shows explicitly the dependence of the critical mass on the chemical composition of the star. 

Following the Stoner's work, S.~Chandrasekhar \cite{chandrasekhar31} \footnote{At the time a 20 years old graduate student coming to Cambridge from India.} pointed out the relevance of describing white dwarfs by using an approach, initiated by E.~A.~Milne \cite{milne30}, of using the mathematical method of the solutions of the Lane-Emden polytropic equations \cite{emdenbook}. The same idea of using the Lane-Emden equations taking into account the special relativistic effects to the equilibrium of stellar matter for a degenerate system of fermions, came independently to L.~D.~Landau \cite{landau32}. Both the Chandrasekhar and Landau treatments were explicit in pointing out the existence of the critical mass
\begin{equation}\label{eq:maxmass}
M^{\rm Ch-L}_{\rm crit} = 2.015 \frac{\sqrt{3 \pi}}{2} \frac{M_{\rm Pl}^3}{\mu^2 m^2_n}\approx 3.09\frac{M_{\rm Pl}^3}{\mu^2 m^2_n} \, ,
\end{equation}
where the first numerical factor on the right hand side of Eq.~(\ref{eq:maxmass}) comes from the boundary condition $-(r^2 d u/dr)_{r=R} = 2.015$ (see last entry of Table 7 on Pag.~80 in \cite{emdenbook}) of the $n=3$ Lane-Emden polytropic equation. Namely for $M > M^{\rm Ch-L}_{\rm crit}$, no equilibrium configuration should exist. 

%This unexpected result created a wave of emotional reactions: 
Landau rejected the idea of the existence of such a critical mass as a ``ridiculous tendency'' \cite{landau32}. Chandrasekhar was confronted by a lively dispute with A.~Eddington on the basic theoretical assumptions he adopted \footnote{The dispute reached such a heated level that Chandrasekhar was confronted with the option either to change field of research or to leave Cambridge. As is well known he chose the second option transferring to Yerkes Observatory near Chicago where he published his results in his classic book \cite{chandrasekharbook}.} (see \cite{wali82} for details). 

Some of the basic assumptions adopted by Chandrasekhar and Landau in their idealized approach were not justified e.g. the treatment of the electron as a free-gas without taking into due account the electromagnetic interactions, as well as the stability of the distribution of the nuclei against the gravitational interaction. It is not surprising that such an approach led to the criticisms of Eddington who had no confidence of the physical foundation of the Chandrasekhar work \footnote{It goes to Eddington credit, at the time Plumian Professor at Cambridge, to have allowed the publication of the Chandrasekhar work although preceded by his own critical considerations \cite{eddington35}.}.
It was unfortunate that the absence of interest of E.~Fermi on the final evolution of stars did not allow Fermi himself to intervene in this contention and solve definitely these well-posed theoretical problems \cite{ruffinibook}. Indeed, we are showing in this article how the solution of the conceptual problems of the white dwarf models, left open for years, can be duly addressed by considering the relativistic Thomas-Fermi model of the compressed atom (see Subsec.~\ref{subsec:reltf} and Sec.~\ref{sec:4}). 

The original work on white dwarfs was motivated by astrophysics and found in astrophysics strong observational support. The issue of the equilibrium of the electron gas and the associated component of nuclei, taking into account the electromagnetic, the gravitational and the weak interactions is a theoretical physics problem, not yet formulated in a correct special and general relativistic context.

One of the earliest alternative approaches to the Chandrasekhar-Landau work was proposed by E.~E.~Salpeter in 1961 \cite{salpeter61}. He followed an idea originally proposed by Y.~I.~Frenkel \cite{frenkel28}: to adopt in the study of white dwarfs the concept of a Wigner-Seitz cell. Salpeter introduced to the lattice model of a point-like nucleus surrounded by a uniform cloud of electrons, corrections due to the non-uniformity of the electron distribution (see Subsec.~\ref{subsec:salpeter} for details). In this way Salpeter \cite{salpeter61} obtained an analytic formula for the total energy in a Wigner-Seitz cell and derived the corresponding equation of state of matter composed by such cells, pointing out explicitly the relevance of the Coulomb interaction. 

The consequences of the Coulomb interactions in the determination of the mass and radius of white dwarfs, was studied in a subsequent paper by T.~Hamada and E.~E.~Salpeter \cite{hamada61} by using the equation of state constructed in \cite{salpeter61}. They found that the critical mass of white dwarfs depends in a nontrivial way on the specific nuclear composition: the critical mass of Chandrasekhar-Landau which depends only on the mass to charge ratio of nuclei $A/Z$, now depends also on the proton number $Z$. 

This fact can be seen from the approximate expression for the critical mass of white dwarfs obtained by Hamada and Salpeter \cite{hamada61} in the ultrarelativistic limit for the electrons
\begin{equation}\label{eq:maxmassHS}
M^{\rm H\&S}_{\rm crit} = 2.015 \frac{\sqrt{3 \pi}}{2} \frac{1}{\mu^2_{\rm eff}} \frac{M_{\rm Pl}^3}{m^2_n}\, ,
\end{equation}
where 
\begin{equation}\label{eq:mueff}
\mu_{\rm eff} = \mu \left(\frac{P_{\rm S}}{P_{\rm Ch}}\right)^{-3/4}\, ,
\end{equation}
being $P_{\rm S}$ the pressure of the Wigner-Seitz cell obtained by Salpeter in \cite{salpeter61} (see Subsec.~\ref{subsec:salpeter}) and $P_{\rm Ch}$ is the pressure of a free-electron fluid used by Chandrasekhar (see Subsec.~\ref{subsec:uniform}). The ratio $P_{\rm S}/P_{\rm Ch}$ is a function of the number of protons $Z$ (see Eq.~(20) in \cite{salpeter61}) and it satisfies $P_{\rm S}/P_{\rm Ch} < 1$. Consequently, the effective molecular weight satisfies $\mu_{\rm eff} > \mu$ and the critical mass of white dwarfs turns to be smaller than the original one obtained by Chandrasekhar-Landau (see Eq.~(\ref{eq:maxmass})).

In the mean time, the problem of the equilibrium gas in a white dwarf taking into account possible global electromagnetic interactions between the nucleus and the electrons was addressed by E.~Olson and M.~Bailyn in \cite{olson75,olson76}. They well summarized the status of the problem: ``\emph{Traditional models for the white dwarf are non-relativistic and electrically neutral ... although an electric field is needed to support the pressureless nuclei against gravitational collapse, the star is treated essentially in terms of only one charge component, where charge neutrality is assumed }''. Their solution to the problem invokes the breakdown of the local charge neutrality and the presence of an overall electric field as a consequence of treating also the nuclei inside the white dwarf as a fluid. They treated the white dwarf matter through a two-fluid model not enforcing local charge neutrality. The closure equation for the Einstein-Maxwell system of equations was there obtained from a minimization procedure of the mass-energy of the configuration. This work was the first pointing out the relevance of the Einstein-Maxwell equations in the description of an astrophysical system by requiring global and non local charge neutrality. As we will show here, this interesting approach does not apply to the case of white dwarfs. It represents, however, a new development in the study of neutron stars (see e.g.~\cite{PLB2011})

An alternative approach to the Salpeter treatment of a compressed atom was reconsidered in \cite{gursky2000} by applying for the first time to white dwarfs a relativistic Thomas-Fermi treatment of the compressed atom introducing a finite size nucleus within a phenomenological description (see also \cite{bertone2000}).

Recently, the study of a compressed atom has been revisited in \cite{2011PhRvC..83d5805R} by extending the global approach of Feynman, Metropolis and Teller \cite{feynman49} taking into account weak interactions. This treatment takes also into account all the Coulomb contributions duly expressed relativistically without the need of any piecewise description. The relativistic Thomas-Fermi model has been solved by imposing in addition to the electromagnetic interaction also the weak equilibrium between neutrons, protons and electrons self-consistently. This presents some conceptual differences with respect to previous approaches and can be used in order both to validate and to establish their limitations.

In this article we apply the considerations presented in \cite{2011PhRvC..83d5805R} of a compressed atom in a Wigner-Seitz cell to the description of non-rotating white dwarfs in general relativity. This approach improves all previous treatments in the following aspects:

\begin{enumerate}
\item In order to warranty self-consistency with a relativistic treatment of the electrons, the point-like assumption of the nucleus is abandoned introducing a finite sized nucleus \cite{2011PhRvC..83d5805R}. We assume for the mass as well as for charge to mass ratio of the nucleus their experimental values instead of using phenomenological descriptions based on the semi-empirical mass-formula of Weizsacker (see e.g.~\cite{gursky2000,bertone2000}).
\item The electron-electron and electron-nucleus Cou\-lomb interaction energy is calculated without any approximation by solving numerically the relativistic Thomas-Fermi equation for selected energy-densities of the system and for each given nuclear composition.
\item The energy-density of the system is calculated taking into account the contributions of the nuclei, of the  Coulomb interactions as well as of the relativistic electrons; the latter being neglected in all previous treatments. This particular contribution turns to be very important at high-densities and in particular for light nuclear compositions e.g.~$^4$He and $^{12}$C.
\item The $\beta$-equilibrium between neutrons, protons, and electrons is also taken into account leading to a self-consistent calculation of the threshold density for triggering the inverse $\beta$-decay of a given nucleus.
\item The structure of the white dwarf configurations is obtained by integrating the general relativity equations of equilibrium.
\item Due to 4) and 5) we are able to determine if the instability point leading to a maximum stable mass of the non-rotating white dwarf is induced by the inverse $\beta$-decay instability of the composing nuclei or by general relativistic effects.
\end{enumerate}

Paradoxically, after all this procedure which takes into account many additional theoretical features generalizing the Chandrasekhar-Landau and the Hamada and Salpeter works, a most simple equation is found to be fulfilled by the equilibrium configuration in a spherically symmetric metric. Assuming the metric 
\begin{equation}\label{eq:metric}
ds^2 = e^{\nu(r)} c^2 dt^2 - e^{\lambda(r)}dr^2 - r^2 d\theta^2 - r^2 \sin^2 \theta d\varphi^2\, ,
\end{equation}
we demonstrate how the entire system of equations describing the equilibrium of white dwarfs, taking into account the weak, the electromagnetic and the gravitational interactions as well as quantum statistics all expressed consistently in a general relativistic approach, is simply given by
\begin{equation}\label{eq:conslaw}
\sqrt{g_{00}} \mu_{\rm ws} = e^{\nu(r)/2}\mu_{\rm ws}(r) = {\rm constant}\, ,
\end{equation}
which links the chemical potential of the Wigner-Seitz cell $\mu_{\rm ws}$, duly solved by considering the relativistic Feynman-Metropolis-Teller model following \cite{2011PhRvC..83d5805R}, to the general relativistic gravitational potential at each point of the configuration. The overall system outside each Wigner-Seitz cell is strictly neutral and no global electric field exists, contrary to the results reported in \cite{olson76}. The same procedure will apply as well to the case of neutron star crusts. 

The article is organized as follows. In Sec.~\ref{sec:2} we summarize the most common approaches used for the description of white dwarfs and neutron star crusts: the uniform approximation for the electron fluid (see e.g.~\cite{chandrasekhar31}); the often called lattice model assuming a point-like nucleus surrounded by a uniform electron cloud (see e.g.~\cite{baym71a}); the generalization of the lattice model due to Salpeter \cite{salpeter61}; the Feynman, Metropolis and Teller approach \cite{feynman49} based on the the non-relativistic Thomas-Fermi model of compressed atoms and, the relativistic generalization of the Feynman-Metropolis-Teller treatment recently formulated in \cite{2011PhRvC..83d5805R}. 

In Sec.~\ref{sec:3} we formulate the general relativistic equations of equilibrium of the system and show how, from the self-consistent definition of chemical potential of the Wigner-Seitz cell and the Einstein equations, comes the equilibrium condition given by Eq.~(\ref{eq:conslaw}). In addition, we obtain the Newtonian and the first-order post-Newtonian equations of equilibrium. 

Finally, we show in Sec.~\ref{sec:4} the new results of the numerical integration of the general relativistic equations of equilibrium and discuss the corrections to the Stoner critical mass $M^{\rm Stoner}_{\rm crit}$, to the Chandrasekhar-Landau mass limit $M^{\rm Ch-L}_{\rm crit}$, as well as to the one of Hamada and Salpeter $M^{\rm H\&S}_{\rm crit}$, obtained when all interactions are fully taken into account through the relativistic Feynman-Metropolis-Teller equation of state \cite{2011PhRvC..83d5805R}.

%%%%%%%%%%%%%%%%%%%%%%%%%%%%%%%%%%%%%%%%%%%%%%%%%%%%%%%%%%%%%%
%%%%%%%%%%%%%%%%%%%%%%%%%%%%%%%%%%%%%%%%%%%%%%%%%%%%%%%%%%%%%%
\section{The Equation of State}\label{sec:2}
%%%%%%%%%%%%%%%%%%%%%%%%%%%%%%%%%%%%%%%%%%%%%%%%%%%%%%%%%%%%%%
%%%%%%%%%%%%%%%%%%%%%%%%%%%%%%%%%%%%%%%%%%%%%%%%%%%%%%%%%%%%%%

There exists a large variety of approaches to model the equation of state of white dwarf matter, each one characterized by a different way of treating or neglecting the Coulomb interaction inside each Wigner-Seitz cell, which we will briefly review here. Particular attention is given to the calculation of the self-consistent chemical potential of the Wigner-Seitz cell $\mu_{\rm ws}$, which plays a very important role in the conservation law (\ref{eq:conslaw}) that we will derive in Sec.~\ref{sec:3}.

%%%%%%%%%%%%%%%%%%%%%%%%%%%%%%%%%%%%%%%%%%%%%%%%%%%%%%%%%%%%%%
\subsection{The uniform approximation}\label{subsec:uniform}
%%%%%%%%%%%%%%%%%%%%%%%%%%%%%%%%%%%%%%%%%%%%%%%%%%%%%%%%%%%%%%

In the uniform approximation used by Chandrasekhar \cite{chandrasekhar31}, the electron distribution as well as the nucleons are assumed to be locally constant and therefore the condition of local charge neutrality
\begin{equation}\label{eq:gncase1}
n_e = \frac{Z}{A_r}n_N\, ,
\end{equation}
where $A_r$ is the average atomic weight of the nucleus, is applied. Here $n_N$ denotes the nucleon number density and $Z$ is the number of protons of the nucleus. The electrons are considered as a fully degenerate free-gas and then described by Fermi-Dirac statistics. Thus, their number density $n_e$ is related to the electron Fermi-momentum $P^F_e$ by 
\begin{equation}\label{eq:nepe}
n_e = \frac{(P^F_e)^3}{3 \pi^2 \hbar^3}\, ,
\end{equation}
and the total electron energy-density and electron pressure are given by 
\begin{eqnarray}
{\cal E}_e &=&  \frac{2}{(2 \pi \hbar)^3} \int_0^{P^F_e} \sqrt{c^2 p^2+m^2_e c^4} 4 \pi p^2 dp \nonumber \\
&=& \frac{m^4_e c^5}{8 \pi^2 \hbar^3}  [x_e \sqrt{1+x^2_e} (1+2 x^2_e)-{\rm arcsinh}(x_e)] \, ,\label{eq:eos1}\\
P_e &=& \frac{1}{3} \frac{2}{(2 \pi \hbar)^3} \int_0^{P^F_e} \frac{c^2 p^2}{\sqrt{c^2 p^2+m^2_e c^4}} 4 \pi p^2 dp \nonumber \\
&=& \frac{m^4_e c^5}{8 \pi^2 \hbar^3}[x_e \sqrt{1+x^2_e}(2x^2_e/3-1)\nonumber \\
&+& {\rm arcsinh}(x_e)] \label{eq:eos2}\, ,
\end{eqnarray}
where we have introduced the dimensionless Fermi momentum $x_e = P^F_e/(m_e c)$ with $m_e$ the electron rest-mass.

The kinetic energy of nucleons is neglected and therefore the pressure is assumed to be only due to electrons. Thus the equation of state can be written as
\begin{eqnarray}\label{eq:Ewscase1}
{\cal E}_{\rm unif} &=& {\cal E}_N + {\cal E}_e \approx \frac{A_r}{Z} M_u c^2 n_e + {\cal E}_e \, ,\label{eq:Ech}\\
P_{\rm unif} &\approx& P_e\, ,\label{eq:Pch}
\end{eqnarray}
where $M_u = 1.6604\times 10^{-24}$ g is the unified atomic mass and ${\cal E}_e$ and $P_e$ are given by Eqs.~(\ref{eq:eos1})--(\ref{eq:eos2}).

Within this approximation, the total self-consistent chemical potential is given by
\begin{equation}\label{eq:muwscase1}
\mu_{\rm unif} = A_r M_u c^2 + Z \mu_e \, ,
\end{equation}
where
\begin{equation}\label{eq:mue}
\mu_e = \frac{{\cal E}_e + P_e}{n_e} = \sqrt{c^2 (P^F_e)^2 + m^2_e c^4}\, ,
\end{equation}
is the electron free-chemical potential. 

As a consequence of this effective approach which does not take into any account the Coulomb interaction, it is obtained an effective one-component electron-nucleon fluid approach where the kinetic pressure is given by electrons of mass $m_e$ and their gravitational contribution is given by an effective mass $(A_r/Z) M_u$ attached to each electron (see e.g.~\cite{landaubook}). This is even more evident when the electron contribution to the energy-density in Eq.~(\ref{eq:Ech}) is neglected and therefore the energy-density is attributed only to the nuclei. Within this approach followed by Chandrasekhar \cite{chandrasekhar31}, the equation of state reduces to
\begin{eqnarray}\label{eq:Ewscase1ch}
{\cal E}_{\rm Ch} &=& \frac{A_r}{Z} M_u c^2 n_e\, ,\label{eq:Ech2}\\
P_{\rm Ch} &=& P_{\rm unif} = P_e\, .\label{eq:Pch2}
\end{eqnarray}

%%%%%%%%%%%%%%%%%%%%%%%%%%%%%%%%%%%%%%%%%%%%%%%%%%%%%%%%%%%%%%
\subsection{The lattice model}
%%%%%%%%%%%%%%%%%%%%%%%%%%%%%%%%%%%%%%%%%%%%%%%%%%%%%%%%%%%%%%

The first correction to the above uniform model, corresponds to abandon the assumption of the electron-nucleon fluid through the so-called ``lattice'' model which introduces the concept of Wigner-Seitz cell: each cell contains a point-like nucleus of charge $+Z e$ with $A$ nucleons surrounded by a uniformly distributed cloud of $Z$ fully-degenerate electrons. The global neutrality of the cell is guaranteed by the condition 
\begin{equation}\label{eq:gncase2}
Z = V_{\rm ws} n_e = \frac{n_e}{n_{\rm ws}}\, ,
\end{equation}  
where $n_{\rm ws}=1/V_{\rm ws}$ is the Wigner-Seitz cell density and $V_{\rm ws} = 4 \pi R^3_{\rm ws}/3$ is the cell volume.

The total energy of the Wigner-Seitz cell is modified by the inclusion of the Coulomb energy, i.e
\begin{equation}\label{eq:Ewscase2}
E_{\rm L} = {\cal E}_{\rm unif} V_{\rm ws} + E_{C}\, , 
\end{equation}
being
\begin{equation}
E_{C} = E_{e-N} + E_{e-e} = -\frac{9}{10} \frac{Z^2 e^2}{R_{\rm ws}}\, ,\label{eq:latticeenergy}
\end{equation}
where ${\cal E}_{\rm unif}$ is given by Eq.~(\ref{eq:Ech}) and $E_{e-N}$ and $E_{e-e}$ are the electron-nucleus and the electron-electron Coulomb energies
\begin{eqnarray}
E_{e-N} &=& - \int_{0}^{R_{\rm ws}} 4 \pi r^2 \left( \frac{Z e}{r} \right) e n_e  dr \nonumber \\
&=& -\frac{3}{2} \frac{Z^2 e^2}{R_{\rm ws}} \, ,\\
E_{e-e} &=& \frac{3}{5} \frac{Z^2 e^2}{R_{\rm ws}} \, .\label{eq:Eeecase2}
\end{eqnarray}

The self-consistent pressure of the Wigner-Seitz cell is then given by
\begin{equation}\label{eq:Pwscase2}
P_{\rm L} = - \frac{\partial E_{\rm L}}{\partial V_{\rm ws}} = P_{\rm unif} + \frac{1}{3} \frac{E_C}{V_{\rm ws}}\, ,
\end{equation}
where $P_{\rm unif}$ is given by Eq.~(\ref{eq:Pch}). It is worth to recall that the point-like assumption of the nucleus is incompatible with a relativistic treatment of the degenerate electron fluid (see \cite{ferreirinho80,ruffini81} for details). Such an inconsistency has been traditionally ignored by applying, within a point-like nucleus model, the relativistic formulas (\ref{eq:eos1}) and (\ref{eq:eos2}) and their corresponding ultrarelativistic limits (see e.g.~\cite{salpeter61}).

The Wigner-Seitz cell chemical potential is in this case
\begin{equation}\label{eq:muwscase2}
\mu_{\rm L} = E_{\rm L} + P_{\rm L} V_{\rm ws} = \mu_{\rm unif} + \frac{4}{3} E_C\, .
\end{equation}

By comparing Eqs.~(\ref{eq:Pch}) and (\ref{eq:Pwscase2}) we can see that the inclusion of the Coulomb interaction results in a decreasing of the pressure of the cell due to the negative lattice energy $E_C$. The same conclusion is achieved for the chemical potential from Eqs.~(\ref{eq:muwscase1}) and (\ref{eq:muwscase2}).

%%%%%%%%%%%%%%%%%%%%%%%%%%%%%%%%%%%%%%%%%%%%%%%%%%%%%%%%%%%%%%
\subsection{Salpeter approach}\label{subsec:salpeter}
%%%%%%%%%%%%%%%%%%%%%%%%%%%%%%%%%%%%%%%%%%%%%%%%%%%%%%%%%%%%%%

A further development to the lattice model came from Salpeter \cite{salpeter61} whom studied the corrections due to the non-uniformity of the electron distribution inside a Wigner-Seitz cell. 

Following the Chandrasekhar \cite{chandrasekhar31} approximation, Salpeter also neglects the electron contribution to the energy-density. Thus, the first term in the Salpeter formula for the energy of the cell comes from the nuclei energy (\ref{eq:Ech2}). The second contribution is given by the Coulomb energy of the lattice model (\ref{eq:latticeenergy}). The third contribution is obtained as follows: the electron density is assumed as $n_e [1+\epsilon(r)]$, where $n_e = 3Z/(4 \pi R^3_{\rm ws})$ is the average electron density as given by Eq.~(\ref{eq:gncase2}), and $\epsilon(r)$ is considered infinitesimal. The Coulomb potential energy is assumed to be the one of the point-like nucleus surrounded by a uniform distribution of electrons, so the correction given by $\epsilon(r)$ on the Coulomb potential is neglected. The electron distribution is then calculated at first-order by expanding the relativistic electron kinetic energy 
\begin{eqnarray}
\epsilon_k &=& \sqrt{[c P^F_e(r)]^2 + m^2_e c^4}-m_e c^2 \nonumber \\
&=& \sqrt{\hbar^2 c^2 (3 \pi^2 n_e)^{2/3}[1+\epsilon(r)]^{2/3} + m^2_e c^4} \nonumber \\
&-& m_e c^2 ,
\end{eqnarray}
about its value in the uniform approximation 
\begin{equation}
\epsilon^{\rm unif}_k = \sqrt{\hbar^2 c^2 (3 \pi^2 n_e)^{2/3} + m^2_e c^4}-m_e c^2\, ,
\end{equation}
considering as infinitesimal the ratio $eV/E^F_e$ between the Coulomb potential energy $eV$ and the electron Fermi energy 
\begin{equation}
E^F_e = \sqrt{[c P^F_e(r)]^2 + m^2_e c^4}-m_e c^2 - e V\, .
\end{equation}

The influence of the Dirac electron-exchange correction \cite{dirac30} on the equation of state was also considered by Salpeter \cite{salpeter61}. However, adopting the general approach of Migdal et al. \cite{migdal77}, it has been shown that these effects are negligible in the relativistic regime \cite{2011PhRvC..83d5805R}. We will then consider here only the major correction of the Salpeter treatment. 

The total energy of the Wigner-Seitz cell is then given by (see \cite{salpeter61} for details)
\begin{equation}\label{eq:Ewscase3}
E_{\rm S} = E_{\rm Ch} + E_{C} + E^{TF}_{S}\, , 
\end{equation}
being
\begin{equation}
E^{TF}_{S} = -\frac{162}{175} \left( \frac{4}{9 \pi} \right)^{2/3} \alpha^2 Z^{7/3} \mu_e\, ,
\end{equation}
where $E_{\rm Ch}={\cal E}_{\rm Ch} V_{\rm ws}$, $E_{C}$ is given by Eq.~(\ref{eq:latticeenergy}), $\mu_e$ is given by Eq.~(\ref{eq:mue}), and $\alpha=e^2/(\hbar c)$ is the fine structure constant.

Correspondingly, the self-consistent pressure of the Wigner-Seitz cell is
\begin{equation}\label{eq:Pwscase3}
P_{\rm S} = P_{\rm L} + P^{S}_{TF}\, ,
\end{equation}
where
\begin{equation}
P^{S}_{TF} = \frac{1}{3} \left( \frac{P^F_e}{\mu_e} \right)^2 \frac{E^{TF}_{S}}{V_{\rm ws}} \, .
\end{equation}

The Wigner-Seitz cell chemical potential can be then written as
\begin{eqnarray}\label{eq:muwscase3}
\mu_{\rm S} = \mu_{\rm L}+ E^{S}_{TF} \left[ 1 + \frac{1}{3} \left( \frac{P^F_e}{\mu_e} \right)^2 \right]\, .
\end{eqnarray}

From Eqs.~(\ref{eq:Pwscase3}) and (\ref{eq:muwscase3}), we see that the inclusion of each additional Coulomb correction results in a further decreasing of the pressure and of the chemical potential of the cell. The Salpeter approach is very interesting in identifying piecewise Coulomb contribution to the total energy, to the total pressure and, to the Wigner-Seitz chemical potential. However, it does not have the full consistency of the global solutions obtained with the Feynman-Metropolis-Teller approach \cite{feynman49} and its generalization to relativistic regimes \cite{2011PhRvC..83d5805R} which we will discuss in detail below.

%%%%%%%%%%%%%%%%%%%%%%%%%%%%%%%%%%%%%%%%%%%%%%%%%%%%%%%%%%%%%%%%%%%%
\subsection{The Feynman-Metropolis-Teller treatment}\label{subsec:nonreltf}
%%%%%%%%%%%%%%%%%%%%%%%%%%%%%%%%%%%%%%%%%%%%%%%%%%%%%%%%%%%%%%%%%%%%

Feynman, Metropolis and Teller \cite{feynman49} showed how to derive the equation of state of matter at high pressures by considering a Thomas-Fermi model confined in a Wigner-Seitz cell of radius $R_{\rm ws}$. 

The Thomas-Fermi equilibrium condition for degenerate non-relativistic electrons in the cell is expressed by 
\begin{eqnarray}\label{eq:nonreleq}
E_e^F = \frac{(P_e^F)^2}{2m_e} - e V = {\rm constant} > 0\, ,
\end{eqnarray}
where $V$ denotes the Coulomb potential and $E_e^F$ denotes the Fermi energy of electrons, which is positive for configurations subjected to external pressure, namely, for compressed cells.

Defining the function $\phi(r)$ by $eV(r)+E_e^F = e^2 Z \phi(r)/r$, and introducing the dimensionless radial coordinate $\eta$ by $r=b\eta$, where $b=(3\pi)^{2/3}(\lambda_e/\alpha)2^{-7/3} Z^{-1/3}$, being $\lambda_{e} = \hbar /(m_e c)$ the electron Compton wavelength; the Poisson equation from which the Coulomb potential $V$ is calculated self-consistently becomes
\begin{eqnarray}\label{eq:nonreltf}
\frac{d^2\phi(\eta)}{d\eta^2}=\frac{\phi(\eta)^{3/2}}{\eta^{1/2}}\, .
\end{eqnarray}
The boundary conditions for Eq.~(\ref{eq:nonreltf}) follow from the point-like structure of the nucleus $\phi(0)=1$ and, from the global neutrality of the Wigner-Seitz cell $\phi(\eta_0) = \eta_0 d\phi/d\eta|_{\eta=\eta_0}$, where $\eta_0$ defines the dimensionless radius of the Wigner-Seitz cell by $\eta_0=R_{\rm ws}/b$.

For each value of the compression, e.g.~$\eta_0$, it corresponds a value of the electron Fermi energy $E^F_e$ and a different solution of Eq.~(\ref{eq:nonreltf}), which determines the self-consistent Coulomb potential energy $e V$ as well as the self-consistent electron distribution inside the cell through
\begin{equation}\label{eq:nenonreltf}
n_e(\eta) = \frac{Z}{4 \pi b^3} \left[\frac{\phi(\eta)}{\eta}\right]^{3/2}\, .
\end{equation}

In the non-relativistic Thomas-Fermi model, the total energy of the Wigner-Seitz cell is given by (see \cite{slater35, feynman49} for details)
\begin{equation}\label{eq:Ewscase4}
E_{\rm ws} = E_N + E^{(e)}_{k} + E_C\, , 
\end{equation}
being
\begin{eqnarray}
E_N &=& M_N(Z,A) c^2\, ,\\
E^{(e)}_{k} &=& \int_{0}^{R_{\rm ws}} 4 \pi r^2 {\cal E}_e[n_e(r)] dr \nonumber \\
&=& \frac{3}{7} \frac{Z^2 e^2}{b} \left[ \frac{4}{5}\eta^{1/2}_0 \phi^{5/2}(\eta_0) - \phi'(0) \right]\, ,\label{eq:Eknonreltf}\\
E_{C} &=& E_{e-N} + E_{e-e} \nonumber \\
&=& -\frac{6}{7} \frac{Z^2 e^2}{b} \left[ \frac{1}{3}\eta^{1/2}_0 \phi^{5/2}(\eta_0) - \phi'(0) \right]\, ,\label{eq:Ecnonreltf}
\end{eqnarray}
where $M_N(Z,A)$ is the nucleus mass, ${\cal E}_e[n_e(r)]$ is given by Eq.~(\ref{eq:eos1}) and $E_{e-N}$ and $E_{e-e}$ are the electron-nucleus Coulomb energy and the electron-electron Coulomb energy, which are given by
\begin{eqnarray}
E_{e-N} &=& - \int_{0}^{R_{\rm ws}} 4 \pi r^2 \left( \frac{Z e}{r} \right) e n_e(r) dr  \, ,\\
E_{e-e} &=& \frac{1}{2} \int_{0}^{R_{\rm ws}} 4 \pi r^2 e n_e(\vec{r}) dr \nonumber \\
&\times& \int_{0}^{R_{\rm ws}} 4 \pi r'^2 \frac{e n_e(\vec{r}')}{|\vec{r}-\vec{r}'|} dr'\, .
\end{eqnarray}

From Eqs.~(\ref{eq:Eknonreltf}) and (\ref{eq:Ecnonreltf}) we recover the well-known relation between the total kinetic energy and the total Coulomb energy in the Thomas-Fermi model \cite{slater35,feynman49}
\begin{equation}\label{eq:enerunifcoulomb}
E^{(e)}_{k} = E^{\rm unif}_k[n_e(R_{\rm ws})] - \frac{1}{2} E_C\, ,
\end{equation}
where $E^{\rm unif}_k [n_e(R_{\rm ws})]$ is the non-relativistic kinetic energy of a uniform electron distribution of density $n_e(R_{\rm ws})$, i.e.
\begin{equation}\label{eq:Eknonrelunif}
E^{\rm unif}_k [n_e(R_{\rm ws})] = \frac{3}{5} Z^* \mu_e(R_{\rm ws})\, ,
\end{equation}
with $Z^*$ defined by
\begin{equation}\label{eq:Znews}
Z^* = V_{\rm ws} n_e (R_{\rm ws})\, ,
\end{equation}
and $\mu_e(R_{\rm ws}) = \hbar^2 [3 \pi^2 n_e(R_{\rm ws})]^{2/3}/(2 m_e)$.

The self-consistent pressure of the Wigner-Seitz cell given by the non-relativistic Thomas-Fermi model is (see \cite{slater35,feynman49} for details)
\begin{equation}\label{eq:Pwscase4}
P_{\rm TF} = \frac{2}{3} \frac{E^{\rm unif}_k [n_e(R_{\rm ws})]}{V_{\rm ws}} \, .
\end{equation}

The pressure of the Thomas-Fermi model (\ref{eq:Pwscase4}) is equal to the pressure of a free-electron distribution of density $n_e(R_{\rm ws})$. Being the electron density inside the cell a decreasing function of the distance from the nucleus, the electron density at the cell boundary, $n_e(R_{\rm ws})$, is smaller than the average electron distribution $3Z/(4 \pi R^3_{\rm ws})$. Then, the pressure given by (\ref{eq:Pwscase4}) is smaller than the one given by the non-relativistic version of Eq.~(\ref{eq:eos2}) of the uniform model of Subsec.~\ref{subsec:uniform}. Such a smaller pressure, although faintfully given by the expression of a free-electron gas, contains in a self-consistent fashion all the Coulomb effects inside the Wigner-Seitz cell.

The chemical potential of the Wigner-Seitz cell of the non-relativistic Thomas-Fermi model can be then written as
\begin{equation}\label{eq:muwscase4}
\mu_{\rm TF} = M_N(Z,A) c^2 + Z^* \mu_e(R_{\rm ws}) + \frac{1}{2} E_C\, ,
\end{equation}
where we have used Eqs.~(\ref{eq:enerunifcoulomb})--(\ref{eq:Znews}).

Integrating by parts the total number of electrons
\begin{equation}
Z = \int_{0}^{R_{\rm ws}} 4 \pi r^2 n_e(r) dr = Z^* + I(R_{\rm ws})\, ,
\end{equation}
where
\begin{equation}
I(R_{\rm ws}) = \int_{0}^{R_{\rm ws}} \frac{4 \pi}{3} r^3 \frac{\partial n_e(r)}{\partial r} dr\, ,
\end{equation}
we can rewrite finally the following semi-analytical expression of the chemical potential (\ref{eq:muwscase4}) of the cell
\begin{eqnarray}\label{eq:muwscase4final}
\mu_{\rm TF} &=& M_N(Z,A) c^2 + Z \mu^{\rm unif}_e \left[ 1 + \frac{I(R_{\rm ws})}{Z} \right]^{2/3} \nonumber \\
&+& \mu^{\rm unif}_e I(R_{\rm ws}) \left[ 1 + \frac{I(R_{\rm ws})}{Z} \right]^{2/3} + \frac{1}{2} E_C\, ,
\end{eqnarray}
where $\mu^{\rm unif}_e$ is the electron free-chemical potential (\ref{eq:mue}) calculated with the average electron density, namely, the electron chemical potential of the uniform approximation. The function $I(R_{\rm ws})$ depends explicitly on the gradient of the electron density, i.e. on the non-uniformity of the electron distribution. 

In the limit of absence of Coulomb interaction both the last term and the function $I(R_{\rm ws})$ in Eq.~(\ref{eq:muwscase4final}) vanish and therefore in this limit $\mu_{\rm TF}$ reduces to
\begin{equation}
\mu_{\rm TF} \to \mu_{\rm unif}\, ,
\end{equation}
where $\mu_{\rm unif}$ is the chemical potential in the uniform approximation given by Eq.~(\ref{eq:muwscase1}).

%%%%%%%%%%%%%%%%%%%%%%%%%%%%%%%%%%%%%%%%%%%%%%%%%%%%%%%%%%%%%%%%%%%%
\subsection{The relativistic Feynman-Metropolis-Teller treatment}\label{subsec:reltf}
%%%%%%%%%%%%%%%%%%%%%%%%%%%%%%%%%%%%%%%%%%%%%%%%%%%%%%%%%%%%%%%%%%%%

We recall now how the above classic Feynman, Metropolis, and Teller treatment of compressed atoms has been recently generalized to relativistic regimes (see \cite{2011PhRvC..83d5805R} for details). One of the main differences in the relativistic generalization of the Thomas-Fermi equation is that, the point-like approximation of the nucleus, must be abandoned since the relativistic equilibrium condition of compressed atoms
\begin{equation}
E_e^F =  \sqrt{c^2 (P_e^F)^2+m_e^2 c^4} - m_e c^2 - e V(r) = {\rm constant} > 0\, ,
\label{eq:efe}
\end{equation} 
would lead to a non-integrable expression for the electron density near the origin (see e.g.\cite{ferreirinho80,ruffini81}).

It is then assumed a constant distribution of protons confined in a radius $R_c$ defined by
\begin{equation} 
R_c = \Delta \lambda_\pi Z ^{1/3}\, , 
\label{eq:pn}
\end{equation}
where $\lambda_{\pi} = \hbar /(m_\pi c)$ is the pion Compton wavelength.  If the system is at nuclear density $\Delta \approx (r_0/\lambda_\pi) (A/Z)^{1/3}$ with $r_0 \approx 1.2$ fm. Thus, in the case of ordinary nuclei (i.e., for $A/Z \approx 2$) we have $\Delta \approx 1$. Consequently, the proton density can be written as
\begin{equation}
n_p(r) = \frac{Z}{\frac{4}{3}\pi R_c^3}\theta(r-R_c) = \frac{3}{4 \pi} \left(\frac{1}{\Delta \lambda_\pi}\right)^3\theta(r-R_c)\, ,
\label{eq:pnx}
\end{equation}
where $\theta(r-R_c)$ denotes the Heaviside function centered at $R_c$. The electron density can be written as
\begin{equation}
n_e(r) = \frac{(P_e^{F})^3}{3\pi^2 \hbar^3} = \frac {1}{3\pi^2 \hbar^3 c^3}\left[\hat V^2(r)+ 2m_e c^2 \hat V(r)\right]^{3/2}\, ,
\label{eq:elnd}
\end{equation}
where $\hat V = e V + E_e^F$ and we have used Eq.~(\ref{eq:efe}).

The overall Coulomb potential satisfies the Poisson equation 
\begin{equation}\label{eq:eposs}
\nabla^2 V(r)= -4\pi e\left[n_p(r)-n_e(r)\right]\, ,
\end{equation}
with the boundary conditions $dV/dr|_{r=R_{\rm ws}}=0$  and  $V(R_{\rm ws})=0$ due to the global charge neutrality of the cell. 

By introducing the dimensionless quantities $ x= r/\lambda_{\pi}$, $ x_c= R_c/\lambda_{\pi}$, $\chi/r = \hat V(r)/(\hbar c)$ and replacing the particle densities (\ref{eq:pnx}) and (\ref{eq:elnd}) into the Poisson equation (\ref{eq:eposs}), it is obtained the relativistic Thomas-Fermi equation \cite{2008pint.conf..207R}
\begin{eqnarray}\label{eq:relTF}
\frac{1}{3x}  \frac {d^2\chi(x)}{d x^2} &=& -\frac{\alpha}{\Delta^3}\theta( x_c- x) \nonumber \\
&+& \frac {4\alpha}{9\pi}\left[\frac {\chi^2(x)}{ x^2} +2\frac{m_e}{m_\pi}\frac{\chi(x)}{x}\right]^{3/2}\, ,  
\end{eqnarray}
which must be integrated subjected to the boundary conditions $\chi(0)=0$, $\chi(x_{\rm ws})\geq 0$ and $d\chi/dx|_{x=x_{\rm ws}} = \chi(x_{\rm ws})/x_{\rm ws}$, where $x_{\rm ws}=R_{\rm ws}/\lambda_{\pi}$.  

The neutron density $n_n(r)$, related to the neutron Fermi momentum $P_n^F=(3\pi^2 \hbar^3 n_n)^{1/3}$, is determined by imposing the condition of beta equilibrium
\begin{eqnarray}
E^F_n &=& \sqrt{c^2 (P_n^F)^2+m^2_n c^4} - m_n c^2 = \sqrt{c^2 (P_p^F)^2+m^2_p c^4} \nonumber \\ 
&-& m_p c^2 + eV(r) + E_e^F\, ,
\label{eq:betaeq}
\end{eqnarray} 
subjected to the baryon number conservation equation
\begin{equation}
A = \int_{0}^{R_c} 4 \pi r^2 [n_p(r) +  n_n(r)] dr\, .
\end{equation}

\begin{figure}[floatfix]
\includegraphics[width=\hsize,clip]{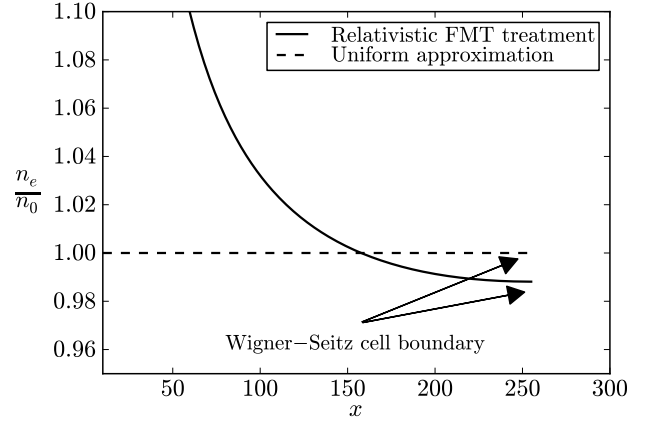}
\caption{The electron number density $n_e$ in units of the average electron number density $n_0 = 3 Z/(4 \pi R_{\rm ws}^3)$ inside a Wigner-Seitz cell of $^{12}$C. The dimensionless radial coordinate is $x=r/\lambda_\pi$ and Wigner-Seitz cell radius is $x_{\rm ws}\approx 255$ corresponding to a density of $\sim 10^8$ g/cm$^3$. The solid curve corresponds to the relativistic Feynman-Metropolis-Teller treatment and the dashed curve to the uniform approximation. The electron distribution for different levels of compression as well as for different nuclear compositions can be found in \cite{2011PhRvC..83d5805R}.}
\label{fig:fmtw}
\end{figure}

In Fig.~\ref{fig:fmtw} we see how the relativistic generalization of the Feynman-Metropolis-Teller treatment leads to electron density distributions markedly different from the constant electron density approximation. The electron distribution is far from being uniform as a result of the solution of Eq.~(\ref{eq:relTF}), which takes into account the electromagnetic interaction between electrons and between the electrons and the finite sized nucleus. Additional details are given in \cite{2011PhRvC..83d5805R}.

V.~S.~Popov et al.~\cite{popov10} have shown how the solution of the relativistic Thomas-Fermi equation (\ref{eq:relTF}) together with the self-consistent implementation of the $\beta$-equilibrium condition (\ref{eq:betaeq}) leads, in the case of zero electron Fermi energy ($E^F_e = 0$), to a theoretical prediction of the $\beta$-equilibrium line, namely a theoretical $Z$-$A$ relation. Within this model the mass to charge ratio $A/Z$ of nuclei is overestimated, e.g. in the case of $^4$He the overestimate is $\sim 3.8\%$, for $^{12}$C $\sim 7.9\%$, for $^{16}$O $\sim 9.52\%$, and for $^{56}$Fe $\sim 13.2\%$. These discrepancies are corrected when the model of the nucleus considered above is improved by explicitly including the effects of strong interactions. This model, however, illustrates how a self-consistent calculation of compressed nuclear matter can be done including electromagnetic, weak, strong as well as special relativistic effects without any approximation. This approach promises to be useful when theoretical predictions are essential, for example in the description of nuclear matter at very high densities, e.g. nuclei close and beyond the neutron drip line. 

The densities in white dwarf interiors are not highly enough to require such theoretical predictions. Therefore, in order to ensure the accuracy of our results we use for $(Z,A)$, needed to solve the relativistic Thomas-Fermi equation (\ref{eq:relTF}), as well as for the nucleus mass $M_N(Z,A)$, their known experimental values. In this way we take into account all the effects of the nuclear interaction.

%In practice, it can be done by using, for instance, the semi-empirical mass-formula of Weizsacker (see e.g.~Segr\`e \cite{segrebook})
%\begin{equation}
%M_N(Z,A) = (A-Z) m_n + Z m_p - \frac{B(Z,A)}{c^2}\, ,
%\end{equation}
%where $B(Z,A)$ is the nuclear binding energy
%\begin{eqnarray}
%B(Z,A) &=& a_{\rm v} A - a_{\rm surf} A^{2/3} - a_{\rm sym} \frac{(A/2-Z)^2}{A} \nonumber \\
%&-& a_{\rm C} \frac{Z^2}{A^{1/3}} - \delta\, ,
%\end{eqnarray}
%whose coefficients are computed from the best-fit of the nuclear masses: $a_{\rm v}=15.67$ MeV, $a_{\rm surf}=17.23$ MeV, $a_{\rm sym} \approx 93.15$ MeV, $a_{\rm C} \approx 0.7$ MeV and $\delta$ is the so-called pairing term accounting for the fact that even-even nuclei are more stable than odd-even and odd-odd nuclei. The $\beta$-stability line, $(\partial M/\partial Z)_A = 0$, gives the $Z$-$A$ relation
%\begin{equation}
%Z = \frac{A}{2 + (2 a_{\rm C}/a_{\rm sym}) A^{2/3}} \approx \frac{A}{2 + 0.015 A^{2/3}}\, .
%\end{equation}

Thus, the total energy of the Wigner-Seitz cell in the present case can be written as
\begin{equation}\label{eq:Ewscase5}
E^{\rm rel}_{\rm FMT} = E_N + E^{(e)}_{k} + E_C\, , 
\end{equation}
being
\begin{eqnarray}
E_N &=& M_N(Z,A) c^2\, ,\label{eq:MNrelTF}\\
E^{(e)}_{k} &=& \int_{0}^{R_{\rm ws}} 4 \pi r^2 ({\cal E}_e - m_e n_e) dr\, ,\label{eq:Ekreltf}\\
E_{C} &=& \frac{1}{2}\int_{R_c}^{R_{\rm ws}} 4 \pi r^2 e [n_p(r) - n_e(r)] V(r) dr \, ,\label{eq:Ecreltf}
\end{eqnarray}
where $M_N(Z,A) = A_r M_u$ is the experimental nucleus mass, e.g. for $^4$He, $^{12}$C, $^{16}$O and $^{56}$Fe we have $A_r=$ 4.003, 12.01, 16.00 and 55.84 respectively. In Eq.~(\ref{eq:Ecreltf}) the integral is evaluated only outside the nucleus (i.e. for $r>R_c$) in order to avoid a double counting with the Coulomb energy of the nucleus already taken into account in the nucleus mass (\ref{eq:MNrelTF}). In order to avoid another double counting we subtract to the electron energy-density ${\cal E}_e$ in Eq.~(\ref{eq:Ekreltf}) the rest-energy density $m_e c^2 n_e$ which is also taken into account in the nucleus mass (\ref{eq:MNrelTF}). 

The total pressure of the Wigner-Seitz cell is given by
\begin{equation}\label{eq:Pwscase5}
P^{\rm rel}_{\rm FMT} = P_e[n_e(R_{\rm ws})]\, ,
\end{equation}
where $P_e[n_e(R_{\rm ws})]$ is the relativistic pressure (\ref{eq:eos2}) computed with the value of the electron density at the boundary of the cell. 

The electron density at the boundary $R_{\rm ws}$ in the relativistic Feynman-Metropolis-Teller treatment is smaller with respect to the one given by the uniform density approximation (see Fig.~\ref{fig:fmtw}). Thus, the relativistic pressure (\ref{eq:Pwscase5}) gives systematically smaller values with respect to the uniform approximation pressure (\ref{eq:eos2}) as well as with respect to the Salpeter pressure (\ref{eq:Pwscase3}).

\begin{figure}
\centering
\includegraphics[width=\columnwidth,clip]{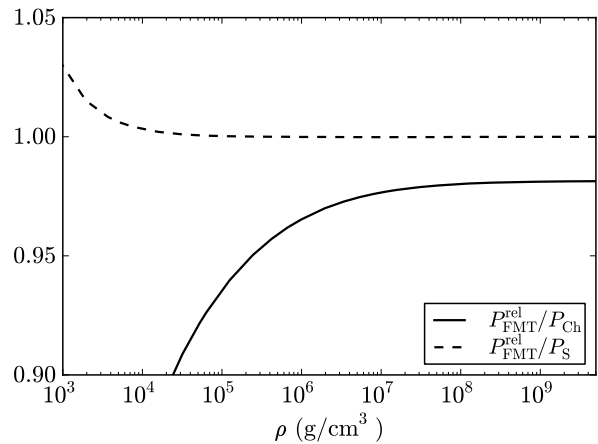}
\caption{Ratio of the pressures in the different treatments as a function of the density for $^{12}$C white dwarfs (see Table \ref{tab:eos}). The solid curve corresponds to the ratio between the relativistic Feynman-Metropolis-Teller pressure $P^{\rm rel}_{\rm FMT}$ given by Eq.~(\ref{eq:Pwscase5}) and the Chandrasekhar pressure $P_{\rm Ch}$ given by Eq.~(\ref{eq:eos2}). The dashed curve corresponds to the ratio between the relativistic Feynman-Metropolis-Teller pressure $P^{\rm rel}_{\rm FMT}$ given by Eq.~(\ref{eq:Pwscase5}) and the Salpeter pressure $P_{\rm S}$ given by Eq.~(\ref{eq:Pwscase3}).}\label{fig:Pratio}
\end{figure}

In Fig.~\ref{fig:Pratio} we show the ratio between the relativistic Feynman-Metropolis-Teller pressure $P^{\rm rel}_{\rm FMT}$ (\ref{eq:Pwscase5}) and the Chandrasekhar pressure $P_{\rm Ch}$ (\ref{eq:eos2}) and the Salpeter pressure $P_{\rm S}$ (\ref{eq:Pwscase3}) in the case of $^{12}$C. It can be seen how $P^{\rm rel}_{\rm FMT}$ is smaller than $P_{\rm Ch}$ for all densities as a consequence of the Coulomb interaction. With respect to the Salpeter case, we have that the ratio $P^{\rm rel}_{\rm FMT}/P_{\rm S}$ approaches unity from below at large densities as one should expect. 

However, at low densities $\lesssim 10^4$--$10^5$ g/cm$^3$, the ratio becomes larger than unity due to the defect of the Salpeter treatment which, in the low density non-relativistic regime, leads to a drastic decrease of the pressure  and even to negative pressures at densities $\lesssim 10^2$ g/cm$^3$ or higher for heavier nuclear compositions e.g.~$^{56}$Fe (see \cite{salpeter61,2011PhRvC..83d5805R} and Table \ref{tab:eos}). This is in contrast with the relativistic Feynman-Metropolis-Teller treatment which matches smoothly the classic Feynman-Metropolis-Teller equation of state in that regime (see \cite{2011PhRvC..83d5805R} for details).

\begin{table}[floatfix]
\begin{center}
\begin{ruledtabular}
\begin{tabular}{c c c c c}
$\rho$ & $P_{\rm Ch}$ & $ P_{\rm S}$ & $P^{\rm rel}_{\rm FMT}$ \\  
\hline
10 & $1.46731 \times 10^{14}$ & $-1.35282 \times 10^{13}$ & $4.54920 \times 10^{14}$ \\
40 & $1.47872 \times 10^{15}$ & $4.60243 \times 10^{14}$ & $7.09818 \times 10^{14}$ \\
70 & $3.75748 \times 10^{15}$ & $1.60860 \times 10^{15}$ & $2.05197 \times 10^{15}$ \\
$10^2$ & $6.80802 \times 10^{15}$ & $3.34940 \times 10^{15}$ & $3.90006 \times 10^{15}$ \\
$10^3$ & $3.15435 \times 10^{17}$ & $2.40646 \times 10^{17}$ & $2.44206 \times 10^{17}$ \\
$10^4$ & $1.45213 \times 10^{19}$ & $1.28976 \times 10^{19}$ & $1.28965 \times 10^{19}$ \\
$10^5$ & $6.50010 \times 10^{20}$ & $6.14494 \times 10^{20}$ & $6.13369 \times 10^{20}$ \\
$10^6$ & $2.62761 \times 10^{22}$ & $2.54932 \times 10^{22}$ & $2.54431 \times 10^{22}$ \\
$10^7$ & $8.46101 \times 10^{23}$ & $8.28899 \times 10^{23}$ & $8.27285 \times 10^{23}$ \\
$10^8$ & $2.15111 \times 10^{25}$ & $2.11375 \times 10^{25}$ & $2.10896 \times 10^{25}$ \\
$10^9$ & $4.86236 \times 10^{26}$ & $4.78170 \times 10^{26}$ & $4.76613 \times 10^{26}$ \\
$10^{10}$ & $1.05977\times 10^{28}$ & $1.04239 \times 10^{28}$ & $1.03668 \times 10^{28}$ \\
\end{tabular}
\end{ruledtabular}
\caption{Equation of state for $^{12}$C within the different treatments. The pressure in the uniform approximation for $\mu=2$ is $P_{\rm Ch}$, the Salpeter pressure is $P_S$ and the relativistic Feynman-Metropolis-Teller pressure is $P^{\rm rel}_{\rm FMT}$. The units for the density are g/cm$^3$ and for the pressure dyn/cm$^2$.}\label{tab:eos}
\end{center}
\end{table}

No analytic expression of the Wigner-Seitz cell chemical potential can be given in this case, so we only write its general expression
\begin{equation}\label{eq:muwscase5}
\mu^{\rm rel}_{\rm FMT} = E^{\rm rel}_{\rm FMT} + P^{\rm rel}_{\rm FMT} V_{\rm ws}\, ,
\end{equation}
where $E^{\rm rel}_{\rm FMT}$ and $P^{\rm rel}_{\rm FMT}$ are given by Eqs.~(\ref{eq:Ewscase5}) and (\ref{eq:Pwscase5}) respectively. The above equation, contrary to the non-relativistic formula (\ref{eq:muwscase4}), in no way can be simplified in terms of its uniform counterparts. However, it is easy to check that, in the limit of no Coulomb interaction $ n_e(R_{\rm ws}) \to 3Z/(4 \pi R^3_{\rm ws})$, $E_C \to 0$, and $E_{k} \to {\cal E}_{\rm Ch} V_{\rm ws}$ and, neglecting the nuclear binding and the proton-neutron mass difference, we finally obtain
\begin{equation}
\mu^{\rm rel}_{\rm FMT} \to \mu_{\rm unif}\, ,
\end{equation}
as it should be expected.

Now we summarize how the equation of state of compressed nuclear matter can be computed in the Salpeter case and in the relativistic Feynman-Metropolis-Teller case, parameterized by the total density of the system:

(i) For a given radius $R_{\rm ws}$ of the Wigner-Seitz cell the relativistic Thomas-Fermi equation (\ref{eq:relTF}) is integrated numerically and the density of the configuration is computed as $\rho = E^{\rm rel}_{\rm FMT}/(c^2 V_{\rm ws})$ where $E^{\rm rel}_{\rm FMT}$ is the energy of the cell given by Eq.~(\ref{eq:Ewscase5}).

(ii) For that value of the density, the radius of the Wigner-Seitz cell in the Salpeter treatment is
\begin{equation}
R_{\rm ws} = \left( \frac{3 A_r M_u}{4 \pi \rho} \right)^{1/3}\, ,
\end{equation}
where Eq.~(\ref{eq:Ech2}) has been used. On the contrary, in the relativistic Feynman-Metropolis-Teller treatment no analytic expression relating Wigner-Seitz cell radius and density can be written. 

(iii) From this Wigner-Seitz cell radius, or equivalently using the value of the density, the electron density in the Salpeter model is computed from the assumption of uniform electron distribution and the charge neutrality condition, i.e.~Eq.~(\ref{eq:Ech2}). In the relativistic Feynman-Metropolis-Teller treatment, the electron number density at the boundary of the Wigner-Seitz cell is, following Eq.~(\ref{eq:elnd}), given by 
\begin{equation}
n^{\rm relFMT}_e = \frac {1}{3\pi^2 \lambda^3_\pi}\left[\frac {\chi^2(x_{\rm ws})}{x^2_{\rm ws}} +2\frac{m_e}{m_\pi}\frac{\chi(x_{\rm ws})}{x_{\rm ws}}\right]^{3/2}\, ,
\end{equation}
where the function $\chi(x)$ is the solution of the relativistic Thomas-Fermi equation (\ref{eq:relTF}).

(iv) Finally, with the knowledge of the electron density at $R_{\rm ws}$, the pressure can be calculated. In the Salpeter approach it is given by Eq.~(\ref{eq:Pwscase3}) while in the relativistic Feynman-Metropolis-Teller case it is given by Eq.~(\ref{eq:Pwscase5}). 

%%%%%%%%%%%%%%%%%%%%%%%%%%%%%%%%%%%%%%%%%%%%%%%%%%%%%%%%%%%%%%
%%%%%%%%%%%%%%%%%%%%%%%%%%%%%%%%%%%%%%%%%%%%%%%%%%%%%%%%%%%%%%
\section{General relativistic equations of equilibrium}\label{sec:3}
%%%%%%%%%%%%%%%%%%%%%%%%%%%%%%%%%%%%%%%%%%%%%%%%%%%%%%%%%%%%%%
%%%%%%%%%%%%%%%%%%%%%%%%%%%%%%%%%%%%%%%%%%%%%%%%%%%%%%%%%%%%%%

Outside each Wigner-Seitz cell the system is electrically neutral, thus no overall electric field exists. Therefore, the above equation of state can be used to calculate the structure of the star through the Einstein equations. Introducing the spherically symmetric metric (\ref{eq:metric}), the Einstein equations can be written in the Tolman-Oppenheimer-Volkoff form \cite{tolman39,oppenheimer39}
\begin{eqnarray}
\frac{d \nu(r)}{dr} &=& \frac{2 G}{c^2} \frac{4 \pi r^3 P(r)/c^2 + M(r)}{r^2 \left[1 - \frac{2 G M(r)}{c^2 r}\right]}\, ,\label{eq:Gab1}\\
\frac{d M(r)}{dr} &=& 4 \pi r^2 \frac{{\cal E}(r)}{c^2}\, ,\label{eq:Gab2}\\
\frac{d P(r)}{dr} &=& - \frac{1}{2} \frac{d \nu(r)}{dr} [{\cal E}(r)+P(r)]\, ,\label{eq:TOV}
\end{eqnarray}
where we have introduced the mass enclosed at the distance $r$ through $e^{\lambda(r)} = 1 - 2 G M(r)/(c^2 r)$, ${\cal E}(r)$ is the energy-density and $P(r)$ is the total pressure.

We turn now to demonstrate how, from Eq.~(\ref{eq:TOV}), it follows the general relativistic equation of equilibrium (\ref{eq:conslaw}), for the self-consistent Wigner-Seitz chemical potential $\mu_{\rm ws}$. The first law of thermodynamics for a zero temperature fluid of $N$ particles, total energy $E$, total volume $V$, total pressure $P = -\partial E/\partial V$, and chemical potential $\mu = \partial E/\partial N$ reads
\begin{equation}\label{eq:thermo1}
d E = - P d V + \mu d N\, ,
\end{equation}
where the differentials denote arbitrary but simultaneous changes in the variables. Since for a system whose surface energy can be neglected with respect to volume energy, the total energy per particle $E/N$ depends only on the particle density $n = N/V$, we can assume $E/N$ as an homogeneous function of first-order in the variables $N$ and $V$ and hence, it follows the well-known thermodynamic relation
\begin{equation}\label{eq:thermo2}
E = -P V + \mu N\, .
\end{equation}

In the case of the Wigner-Seitz cells, Eq.~(\ref{eq:thermo2}) reads
\begin{equation}\label{eq:thermo3}
E_{\rm ws} = -P_{\rm ws} V_{\rm ws} + \mu_{\rm ws}\, ,
\end{equation}
where we have introduced the fact that the Wigner-Seitz cells are the building blocks of the configuration and therefore we must put in Eq.~(\ref{eq:thermo2}) $N_{\rm ws} = 1$. Through the entire article we have used Eq.~(\ref{eq:thermo3}) to obtain from the knowns energy and pressure, the Wigner-Seitz cell chemical potential (see e.g. Eqs.~(\ref{eq:muwscase1}) and (\ref{eq:muwscase2})). From Eqs.~(\ref{eq:thermo1}) and (\ref{eq:thermo2}) we obtain the so-called Gibbs-Duhem relation
\begin{equation}\label{eq:gibbsduhem}
d P = n d \mu\, .
\end{equation}

In a white dwarf the pressure $P$ and the chemical potential $\mu$ are decreasing functions of the distance from the origin. Thus, the differentials in the above equations can be assumed as the gradients of the variables which, in the present spherically symmetric case, become just derivatives with respect to the radial coordinate $r$. From Eq.~(\ref{eq:gibbsduhem}) it follows the relation
\begin{equation}\label{eq:gibbsduhemws}
\frac{dP_{\rm ws}}{dr} = n_{\rm ws} \frac{d\mu_{\rm ws}}{dr}\, .
\end{equation}
From Eqs.~(\ref{eq:TOV}), (\ref{eq:thermo3}) and (\ref{eq:gibbsduhemws}) we obtain
\begin{equation}
n_{\rm ws}(r) \frac{d\mu_{\rm ws}(r)}{dr} = - \frac{1}{2} \frac{d \nu(r)}{dr} n_{\rm ws}(r)\mu_{\rm ws}(r) \, ,\label{eq:TOV2}
\end{equation}
which can be straightforwardly integrated to obtain the first integral
\begin{equation}\label{eq:conslaw2}
e^{\nu(r)/2}\mu_{\rm ws}(r) = {\rm constant}\, .
\end{equation}

The above equilibrium condition is general and it also applies for non-zero temperature configurations ( see e.g.~\cite{klein49}). In such a case, it can be shown that in addition to the equilibrium condition (\ref{eq:conslaw2}) the temperature of the system satisfies the Tolman isothermality condition $e^{\nu(r)/2}T(r) = $ constant \cite{1930PhRv...35..904T,1930PhRv...36.1791T}.

%%%%%%%%%%%%%%%%%%%%%%%%%%%%%%%%%%%%%%%%%%%%%%%%%%%%%%%%%%%%
\subsection{The weak-field non-relativistic limit}
%%%%%%%%%%%%%%%%%%%%%%%%%%%%%%%%%%%%%%%%%%%%%%%%%%%%%%%%%%%%

In the weak-field limit we have $e^{\nu/2} \approx 1 + \Phi$, where the Newtonian gravitational potential has been defined by $\Phi (r) = \nu(r)/2$. In the non-relativistic mechanics limit $c\to \infty$, the chemical potential $\mu_{\rm ws} \to \tilde{\mu}_{\rm ws} + M_{\rm ws} c^2$, where $\tilde{\mu}_{\rm ws}$ denotes the non-relativistic free-chemical potential of the Wigner-Seitz cell and $M_{\rm ws}$ is the rest-mass of the Wigner-Seitz cell, namely, the rest-mass of the nucleus plus the rest-mass of the electrons. Applying these considerations to Eq.~(\ref{eq:conslaw2}) we obtain
\begin{equation}
e^{\nu/2} \mu_{\rm ws} \approx M_{\rm ws} c^2 + \tilde{\mu}_{\rm ws} + M_{\rm ws} \Phi = {\rm constant}\, .
\end{equation}
Absorbing the Wigner-Seitz rest-mass energy $M_{\rm ws} c^2$ in the constant on the right-hand-side we obtain
\begin{equation}\label{eq:eqnonrel}
\tilde{\mu}_{\rm ws} + M_{\rm ws} \Phi = {\rm constant}\, .
\end{equation}

In the weak-field non-relativistic limit, the Einstein equations (\ref{eq:Gab1})--(\ref{eq:TOV}) reduce to
\begin{eqnarray}
\frac{d \Phi(r)}{dr} &=& \frac{G M(r)}{r^2}\, ,\label{eq:Gablimit1}\\
\frac{d M(r)}{dr} &=& 4 \pi r^2 \rho(r)\, ,\label{eq:Gablimit2}\\
\frac{d P(r)}{dr} &=& - \frac{G M(r)}{r^2} \rho(r)\, ,\label{eq:Gablimit3}
\end{eqnarray}
where $\rho(r)$ denotes the rest-mass density. The Eqs.~(\ref{eq:Gablimit1})--(\ref{eq:Gablimit2}) can be combined to obtain the gravitational Poisson equation
\begin{equation}\label{eq:poisson}
\frac{d^2 \Phi(r)}{dr^2} +\frac{2}{r}\frac{d\Phi(r)}{dr} = 4 \pi G \rho(r)\, .
\end{equation}

In the uniform approximation (see Subsec.~\ref{subsec:uniform}), the equilibrium condition given by Eq.~(\ref{eq:eqnonrel}) reads
\begin{equation}\label{eq:eleceqnonrel}
\tilde{\mu}_e + \frac{A_r}{Z} M_u \Phi= {\rm constant}\, ,
\end{equation}
where we have neglected the electron rest-mass with respect to the nucleus rest-mass and we have divided the equation by the total number of electrons $Z$. This equilibrium equation is the classical condition of thermodynamic equilibrium assumed for non-relativistic white dwarf models (see e.g.~\cite{landaubook} for details).

Introducing the above equilibrium condition (\ref{eq:eleceqnonrel}) into Eq.~(\ref{eq:poisson}), and using the relation between the non-relativistic electron chemical potential and the particle density $n_e = (2 m_e)^{3/2} \tilde{\mu}^{3/2}_e/(3 \pi^2 \hbar^3)$, we obtain
\begin{equation}\label{eq:newtonianeq}
\frac{d^2 \tilde{\mu}_e(r)}{dr^2} +\frac{2}{r}\frac{d\tilde{\mu}_e(r)}{dr} = - \frac{2^{7/3} m^{3/2}_e (A_r/Z)^2 m^2_N G}{3 \pi \hbar^3} \tilde{\mu}^{3/2}_e(r)\, ,
\end{equation}
which is the correct equation governing the equilibrium of white dwarfs within Newtonian gravitational theory \citep{landaubook}. It is remarkable that the equation of equilibrium (\ref{eq:newtonianeq}), obtained from the correct application of the Newtonian limit, does not coincide with the equation given by \cite{chandrasekhar31,chandrasekhar31a, chandrasekhar35, chandrasekharbook}, which, as correctly pointed out by \cite{eddington35}, is a mixture of both relativistic and non-relativistic approaches. Indeed, the consistent relativistic equations should be Eq.~(\ref{eq:conslaw2}). Therefore a dual relativistic and non-relativistic equation of state was used by Chandrasekhar. The pressure on the left-hand-side of Eq.~(\ref{eq:Gablimit3}) is taken to be given by relativistic electrons while, the term on the right-hand-side of Eq.~(\ref{eq:Gablimit2}) and (\ref{eq:Gablimit3}) (or the source of Eq.~(\ref{eq:poisson})), is taken to be the rest-mass density of the system instead of the total relativistic energy-density. Such a procedure is equivalent to take the chemical potential in Eq.~(\ref{eq:eqnonrel}) as a relativistic quantity. As we have seen, this is inconsistent with the weak-field non-relativistic limit of the general relativistic equations.

%%%%%%%%%%%%%%%%%%%%%%%%%%%%%%%%%%%%%%%%%%%%%%%%%%%%%%%%%%%%
\subsection{The Post-Newtonian limit}
%%%%%%%%%%%%%%%%%%%%%%%%%%%%%%%%%%%%%%%%%%%%%%%%%%%%%%%%%%%%

Although quantitatively justifiable (see next section), the Chandrasekhar approach was strongly criticized by Eddington because it was conceptually unjustified. Indeed, if one were to treat the problem of white dwarfs approximately without going to the sophistications of general relativity, but including the effects of relativistic mechanics, one should use at least the equations in the post-Newtonian limit. The first-order post-Newtonian expansion of the Einstein equations (\ref{eq:Gab1})--(\ref{eq:TOV}) in powers of $P/{\cal E}$ and $G M/(c^2 r)$ leads to the equilibrium equations \citep{ciufolini83}
\begin{eqnarray}
\frac{d \Phi(r)}{dr} &=& -\frac{1}{{\cal E}(r)} \left[ 1 - \frac{P(r)}{{\cal E}(r)}\right] \frac{d P(r)}{dr}\, ,\label{eq:GabPN1}\\
\frac{d M(r)}{dr} &=& 4 \pi r^2 \frac{{\cal E}(r)}{c^2}\, ,\label{eq:GabPN2}\\
\frac{d P(r)}{dr} &=& -\frac{G M(r)}{r^2}\frac{{\cal E}(r)}{c^2} \bigg[ 1 + \frac{P(r)}{{\cal E}(r)} + \frac{4 \pi r^3 P(r)}{M(r) c^2}\nonumber \\
&+& \frac{2 G M(r)}{c^2 r} \bigg]\, ,\label{eq:GabPN3}
\end{eqnarray}
where Eq.~(\ref{eq:GabPN3}) is the post-Newtonian version of the Tolman-Oppenheimer-Volkoff equation (\ref{eq:TOV}).

Replacing Eq.~(\ref{eq:gibbsduhemws}) into Eq.~(\ref{eq:GabPN1}) we obtain
\begin{equation}\label{eq:PNeq1}
\left[ 1 - \frac{P(r)}{{\cal E}(r)}\right]\frac{d \mu_{\rm ws}(r)}{dr} + \frac{{\cal E}(r)/c^2}{n_{\rm ws}(r)} \frac{d \Phi(r)}{dr} = 0\, .
\end{equation}
It is convenient to split the energy-density as ${\cal E} = c^2 \rho + U$, where $\rho = M_{\rm ws} n_{\rm ws}$ is the rest-energy density and $U$ the internal energy-density. Thus, Eq.~(\ref{eq:PNeq1}) becomes
\begin{eqnarray}\label{eq:PNeq2}
\frac{d \mu_{\rm ws}(r)}{dr} &+& M_{\rm ws} \frac{d \Phi(r)}{dr} - \frac{P(r)}{{\cal E}(r)}\frac{d \mu_{\rm ws}(r)}{dr} \nonumber \\
&+& \frac{U/c^2}{n_{\rm ws}(r)} \frac{d \Phi(r)}{dr} = 0\, ,
\end{eqnarray}
which is the differential post-Newtonian version of the equilibrium equation (\ref{eq:conslaw2}) and where the post-Newtonian corrections of equilibrium can be clearly seen. Applying the non-relativistic limit $c \to \infty$ to Eq.~(\ref{eq:PNeq2}): $ P/{\cal E} \to 0$, $U/c^2 \to 0$, and $\mu_{\rm ws} \to M_{\rm ws} c^2 + \tilde{\mu}_{\rm ws}$, we recover the Newtonian equation of equilibrium (\ref{eq:eqnonrel}).

%%%%%%%%%%%%%%%%%%%%%%%%%%%%%%%%%%%%%%%%%%%%%%%%%%%%%%%%%%%%%%
%%%%%%%%%%%%%%%%%%%%%%%%%%%%%%%%%%%%%%%%%%%%%%%%%%%%%%%%%%%%%%
\section{Mass and radius of general relativistic stable white dwarfs}\label{sec:4}
%%%%%%%%%%%%%%%%%%%%%%%%%%%%%%%%%%%%%%%%%%%%%%%%%%%%%%%%%%%%%%
%%%%%%%%%%%%%%%%%%%%%%%%%%%%%%%%%%%%%%%%%%%%%%%%%%%%%%%%%%%%%%
 
%%%%%%%%%%%%%%%%%%%%%%%%%%%%%%%%%%%%%%%%%%%%%%%%%%%%%%%%%%%%%%
\subsection{Inverse $\beta$-decay instability}\label{subsec:betadecay}
%%%%%%%%%%%%%%%%%%%%%%%%%%%%%%%%%%%%%%%%%%%%%%%%%%%%%%%%%%%%%%
  
It is known that white dwarfs may become unstable against the inverse $\beta$-decay process $(Z,A)\to (Z-1,A)$ through the capture of energetic electrons (see e.g.~\cite{hund36,landau38,zeldovich58,harrison58}). In order to trigger such a process, the electron Fermi energy must be larger than the mass difference between the initial nucleus $(Z,A)$ and the final nucleus $(Z-1,A)$. We denote this threshold energy as $\epsilon^\beta_Z$. Usually it is satisfied $\epsilon^\beta_{Z-1} < \epsilon^\beta_Z$ and therefore the initial nucleus undergoes two successive decays, i.e. $(Z,A)\to (Z-1,A)\to (Z-2,A)$ (see e.g.~\cite{salpeter61,shapirobook}). Some of the possible decay channels in white dwarfs with the corresponding known experimental threshold energies $\epsilon^\beta_Z$ are listed in Table \ref{tab:betadecay}. The electrons in the white dwarf may eventually reach the threshold energy to trigger a given decay at some critical density $\rho^{\beta}_{\rm crit}$. Configurations with $\rho > \rho^{\beta}_{\rm crit}$ become unstable (see \cite{harrison58,salpeter61} for details). 

Within the uniform approximation, e.g.~in the case of the Salpeter equation of state \cite{salpeter61}, the critical density for the onset of inverse $\beta$-decay is given by
\begin{equation}\label{eq:betaunif}
\rho^{\beta,\rm unif}_{\rm crit} = \frac{A_r}{Z}\frac{M_u}{3 \pi^2 \hbar^3 c^3} [(\epsilon^\beta_Z)^2 + 2 m_e c^2 \epsilon^\beta_Z]^{3/2}\, ,
\end{equation}
where Eq.~(\ref{eq:Ech2}) has been used.

Because the computation of the electron Fermi energy within the relativistic Feynman-Metropolis-Teller approach \cite{2011PhRvC..83d5805R} involves the numerical integration of the relativistic Thomas-Fermi equation (\ref{eq:relTF}), no analytic expression for $\rho^{\beta}_{\rm crit}$ can be found in this case. The critical density $\rho^{\beta,\rm relFMT}_{\rm crit}$ is then obtained numerically by looking for the density at which the electron Fermi energy (\ref{eq:efe}) equals $\epsilon^\beta_Z$. 

In Table \ref{tab:betadecay} we show, correspondingly to each threshold energy $\epsilon^\beta_Z$, the critical density both in the Salpeter case $\rho^{\beta,\rm unif}_{\rm crit}$ given by Eq.~(\ref{eq:betaunif}) and in the relativistic Feynman-Metropolis-Teller case $\rho^{\beta,\rm relFMT}_{\rm crit}$. It can be seen that $\rho^{\beta,\rm relFMT}_{\rm crit} > \rho^{\beta,\rm unif}_{\rm crit}$ as one should expect from the fact that, for a given density, the electron density at the Wigner-Seitz cell boundary satisfies $n^{\rm relFMT}_e < n^{\rm unif}_e$. This means that, in order to reach a given energy, the electrons within the relativistic Feynman-Metropolis-Teller approach must be subjected to a larger density with respect to the one given by the approximated Salpeter analytic formula (\ref{eq:betaunif}).

\begin{table}[floatfix]
\begin{center}
\begin{ruledtabular}
\begin{tabular}{c c c c c}
Decay & $\epsilon^\beta_Z$ & $\rho^{\beta,\rm relFMT}_{\rm crit}$ & $\rho^{\beta,\rm unif}_{\rm crit}$\\
\hline 
$^{4}$He $\to ^3$ H + $n \to 4 n$ & 20.596 & $1.39\times 10^{11}$ & $1.37\times 10^{11}$\\
$^{12}$C $\to ^{12}$B $\to ^{12}$Be & 13.370 & $3.97\times 10^{10}$ & $3.88\times 10^{10}$\\
$^{16}$O $\to ^{16}$N $\to ^{16}$C& 10.419 & $1.94\times 10^{10}$ & $1.89\times 10^{10}$\\
$^{56}$Fe $\to ^{56}$Mn $\to ^{56}$Cr& 3.695 & $1.18\times 10^{9}$ & $1.14\times 10^{9}$
\end{tabular}
\end{ruledtabular}
\caption{Onset of inverse beta decay instability for $^{4}$He, $^{12}$C, $^{16}$O and $^{56}$Fe. The experimental inverse $\beta$-decay energies $\epsilon^\beta_Z$ are given in MeV and they have been taken from Table 1 of \cite{audi03}. The corresponding critical density for the uniform electron density model, $\rho^{\beta, \rm unif}_{\rm crit}$ given by Eq.~(\ref{eq:betaunif}), is given in g/cm$^3$ as well as the critical density $\rho^{\beta, \rm relFMT}_{\rm crit}$ for the relativistic Feynman-Metropolis-Teller case. The numerical values of $\epsilon^\beta_Z$ are taken from \cite{1977ADNDT..19..175W}, see also \cite{shapirobook}}.\label{tab:betadecay}
\end{center}
\end{table}

%%%%%%%%%%%%%%%%%%%%%%%%%%%%%%%%%%%%%%%%%%%%%%%%%%%%%%%%%%%%%%
\subsection{General relativistic instability}\label{subsec:grinstability}
%%%%%%%%%%%%%%%%%%%%%%%%%%%%%%%%%%%%%%%%%%%%%%%%%%%%%%%%%%%%%%

The concept of the critical mass has played a major role in the theory of stellar evolution. For Newtonian white dwarfs the critical mass is reached asymptotically at infinite central densities of the object. One of the most important general relativistic effects is to shift this critical point to some finite density $\rho^{\rm GR}_{\rm crit}$.

This general relativistic effect is an additional source of instability with respect to the already discussed instability due to the onset of inverse $\beta$-decay which, contrary to the present general relativistic one, applies also in the Newtonian case by shifting the maximum mass of Newtonian white dwarfs to finite densities (see e.g.~\cite{harrison58}). 

%%%%%%%%%%%%%%%%%%%%%%%%%%%%%%%%%%%%%%%%%%%%%%%%%%%%%%%%%%%%%%
\subsection{Numerical results}\label{subsec:results}
%%%%%%%%%%%%%%%%%%%%%%%%%%%%%%%%%%%%%%%%%%%%%%%%%%%%%%%%%%%%%%

\begin{figure*}
\centering
\begin{tabular}{lr}
\includegraphics[width=\columnwidth,clip]{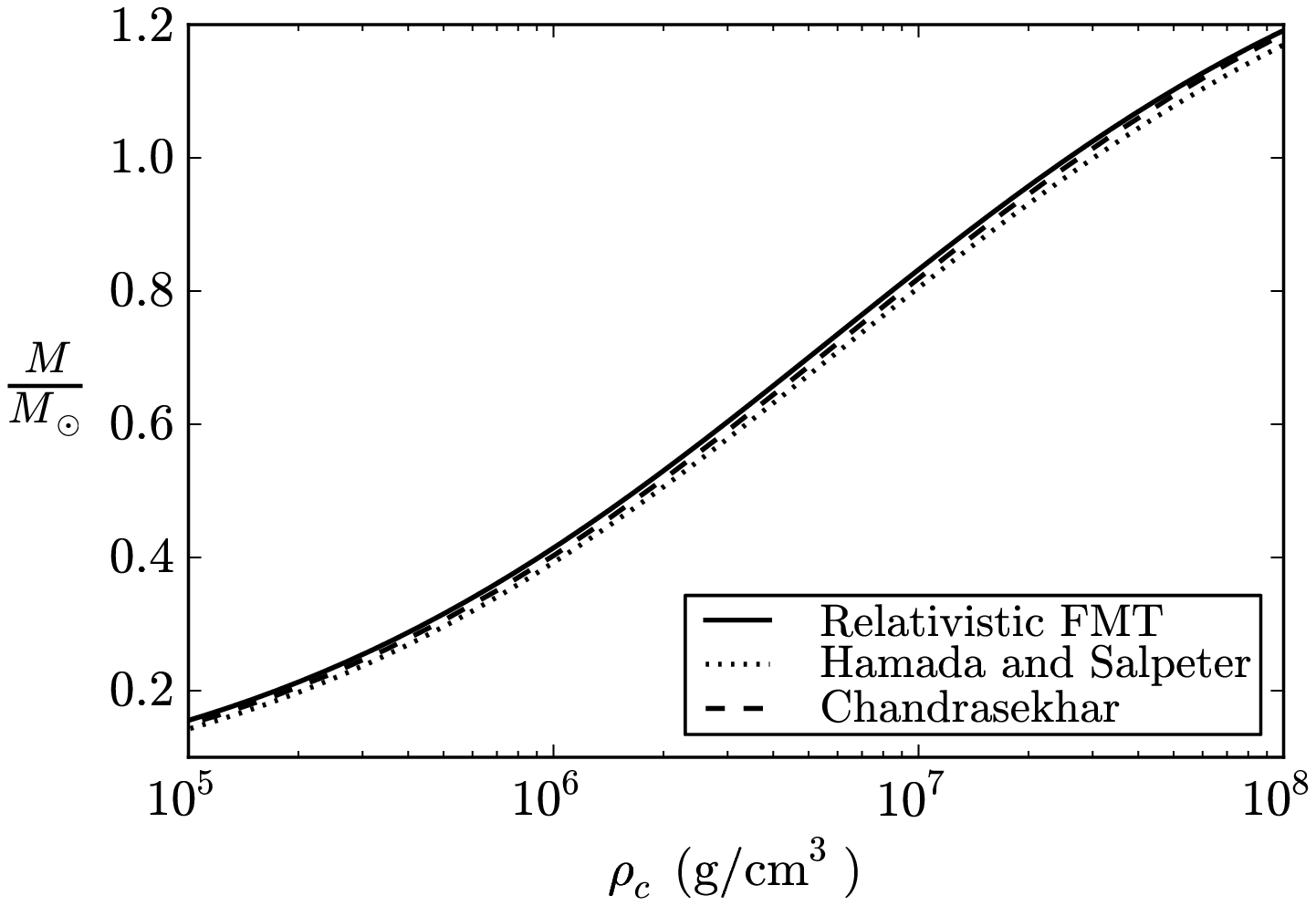} & \includegraphics[width=\columnwidth,clip]{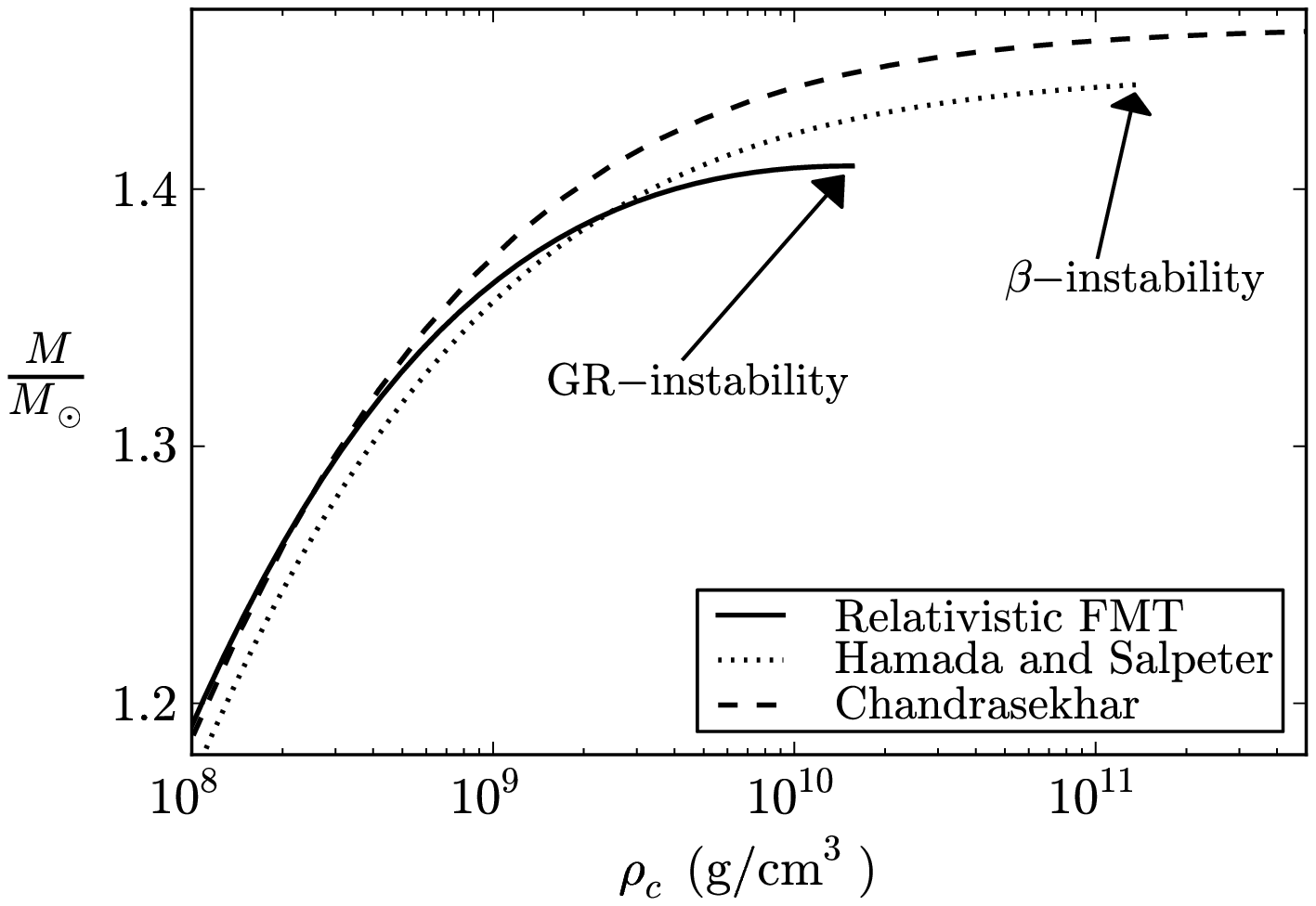}
\end{tabular}
\caption{Mass in solar masses as a function of the central density in the range (left panel) $10^5$--$10^8$ g/cm$^3$ and in the range (right panel) $10^8$--$5\times 10^{11}$ g/cm$^3$ for $^{4}$He white dwarfs. The solid curve corresponds to the present work, the dotted curves are the Newtonian configurations of Hamada and Salpeter and the dashed curve are the Newtonian configurations of Chandrasekhar.}\label{fig:MrhoHe}
\end{figure*}

\begin{figure*}
\centering
\begin{tabular}{lr}
\includegraphics[width=\columnwidth,clip]{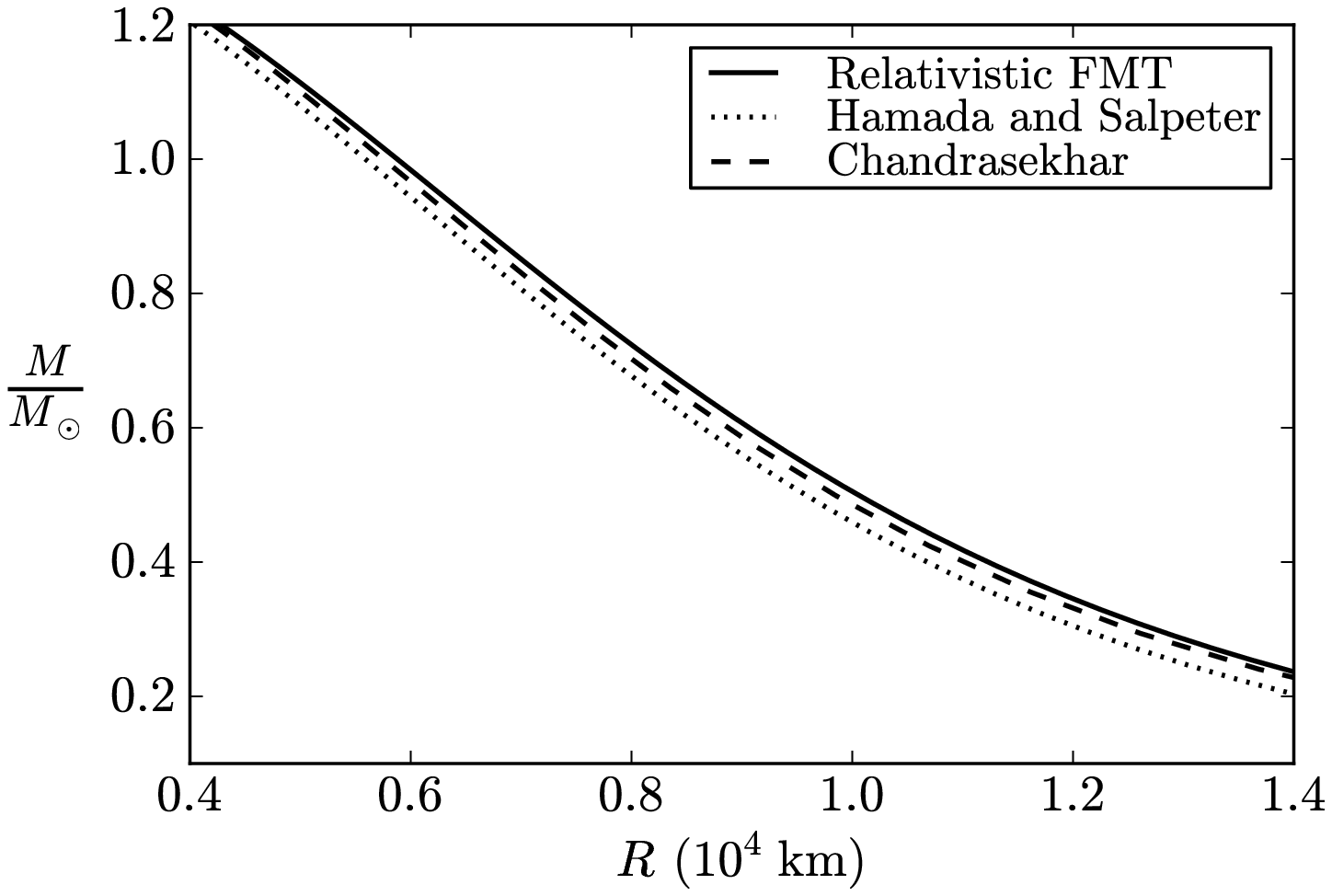} & \includegraphics[width=\columnwidth,clip]{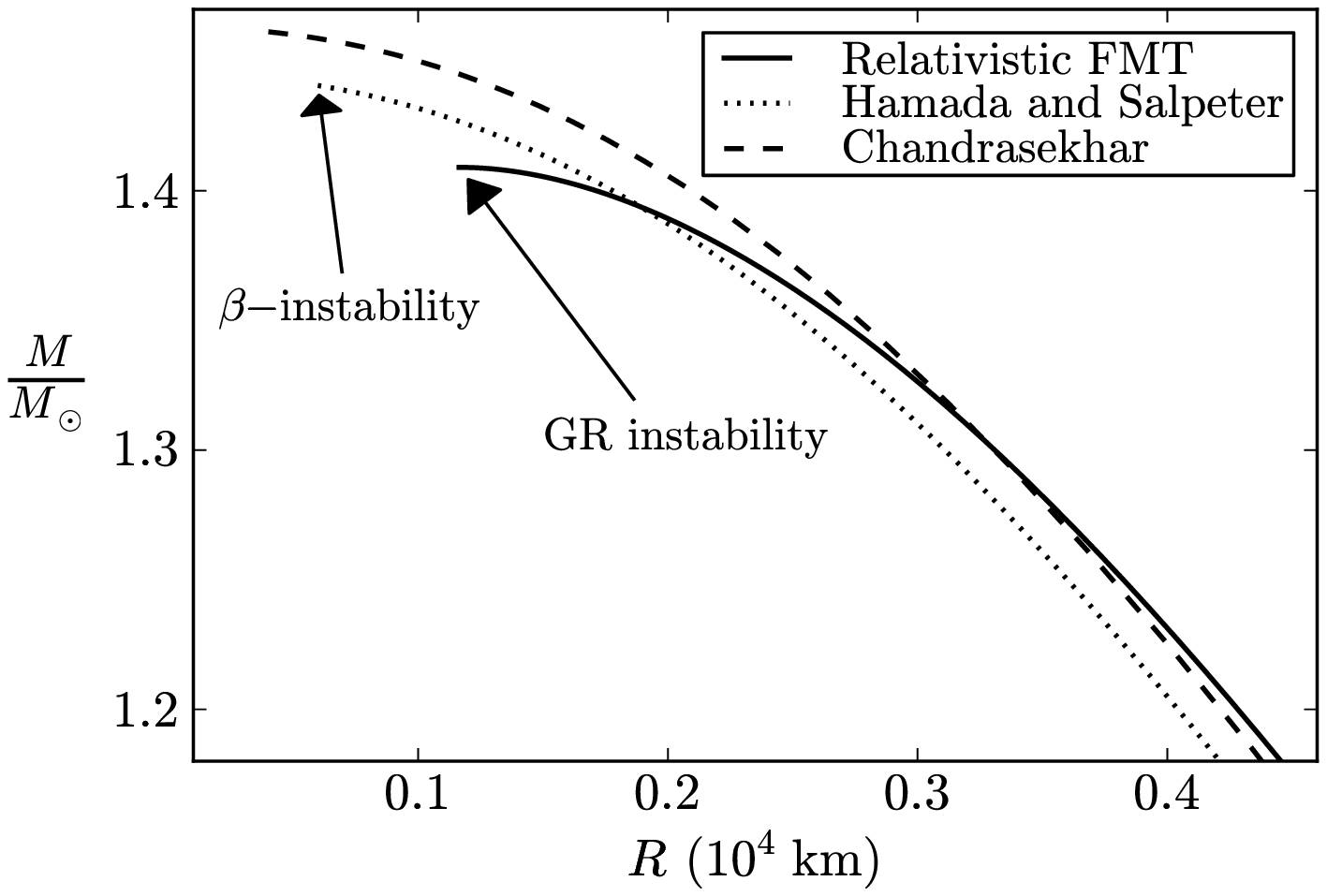}
\end{tabular}
\caption{Mass in solar masses as a function of the radius in units of $10^4$ km for $^{4}$He white dwarfs. The left and right panels show the configurations for the same range of central densities of the corresponding panels of Fig.~\ref{fig:MrhoHe}. The solid curve corresponds to the present work, the dotted curves are the Newtonian configurations of Hamada and Salpeter and the dashed curve are the Newtonian configurations of Chandrasekhar.}\label{fig:MRHe}
\end{figure*}

\begin{figure*}
\centering
\begin{tabular}{lr}
\includegraphics[width=\columnwidth,clip]{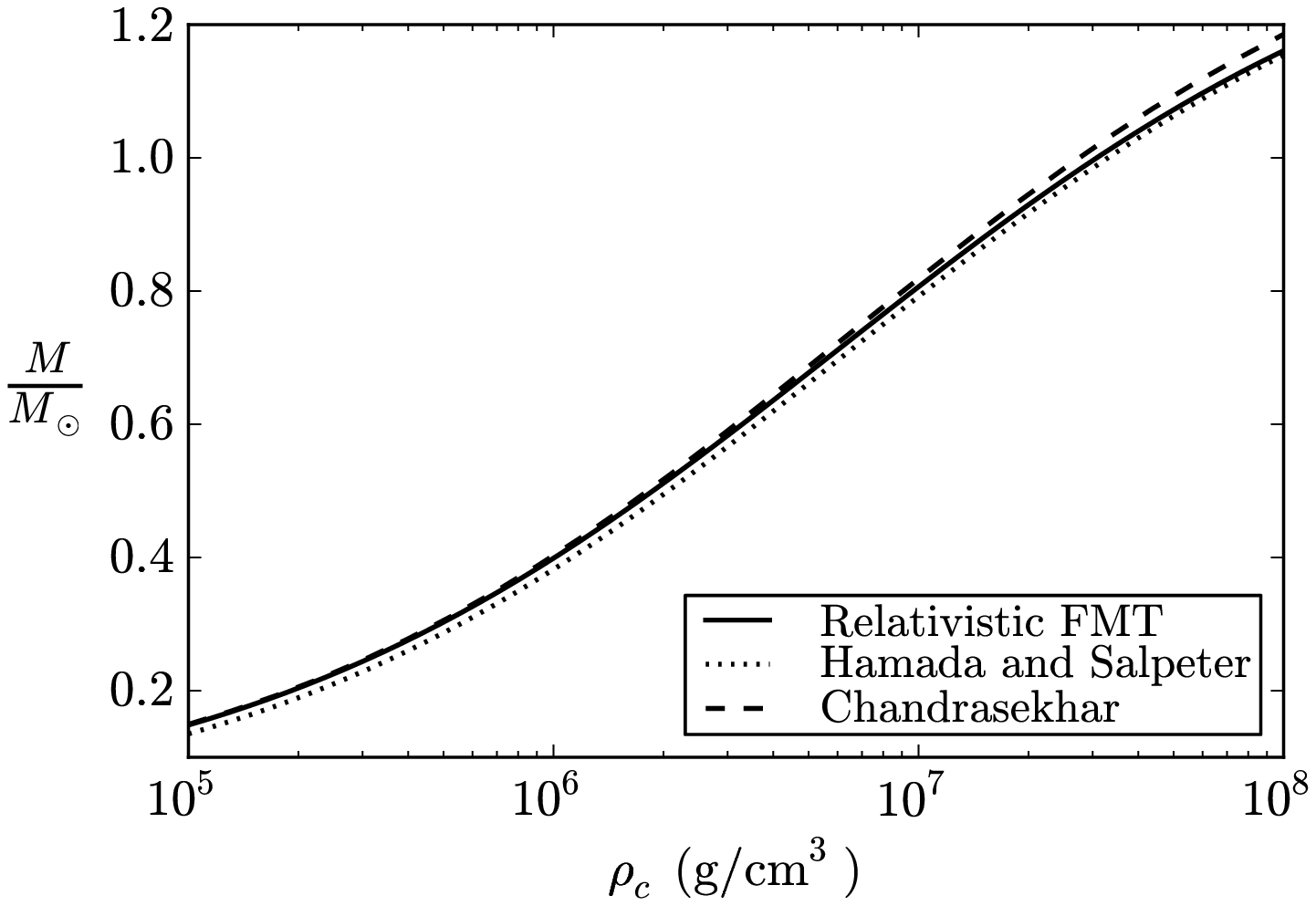} & \includegraphics[width=\columnwidth,clip]{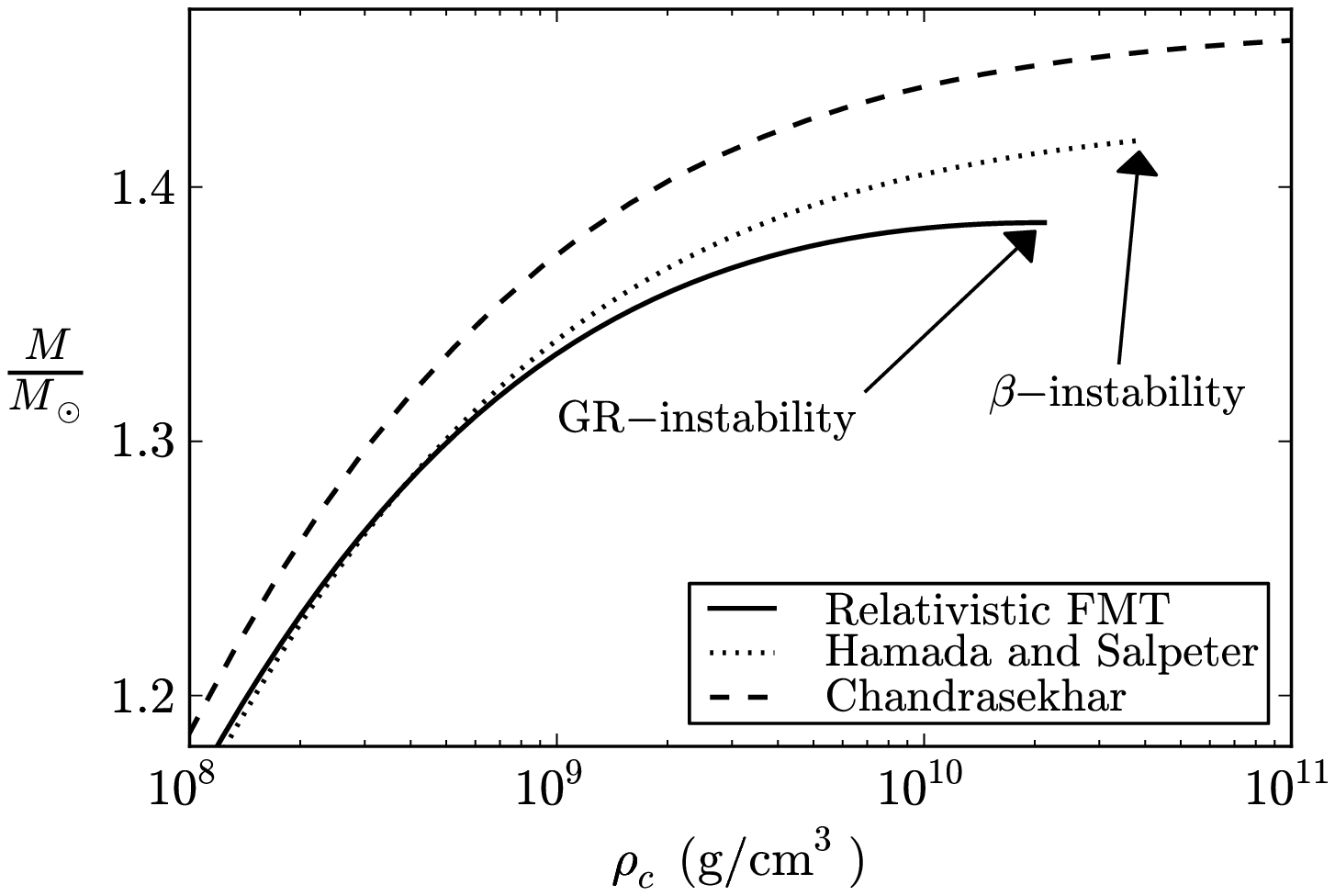}
\end{tabular}
\caption{Mass in solar masses as a function of the central density in the range (left panel) $10^5$--$10^8$ g/cm$^3$ and in the range (right panel) $10^8$--$10^{11}$ g/cm$^3$ for $^{12}$C white dwarfs. The solid curve corresponds to the present work, the dotted curves are the Newtonian configurations of Hamada and Salpeter and the dashed curve are the Newtonian configurations of Chandrasekhar.}\label{fig:MrhoC}
\end{figure*}

\begin{figure*}
\centering
\begin{tabular}{lr}
\includegraphics[width=\columnwidth,clip]{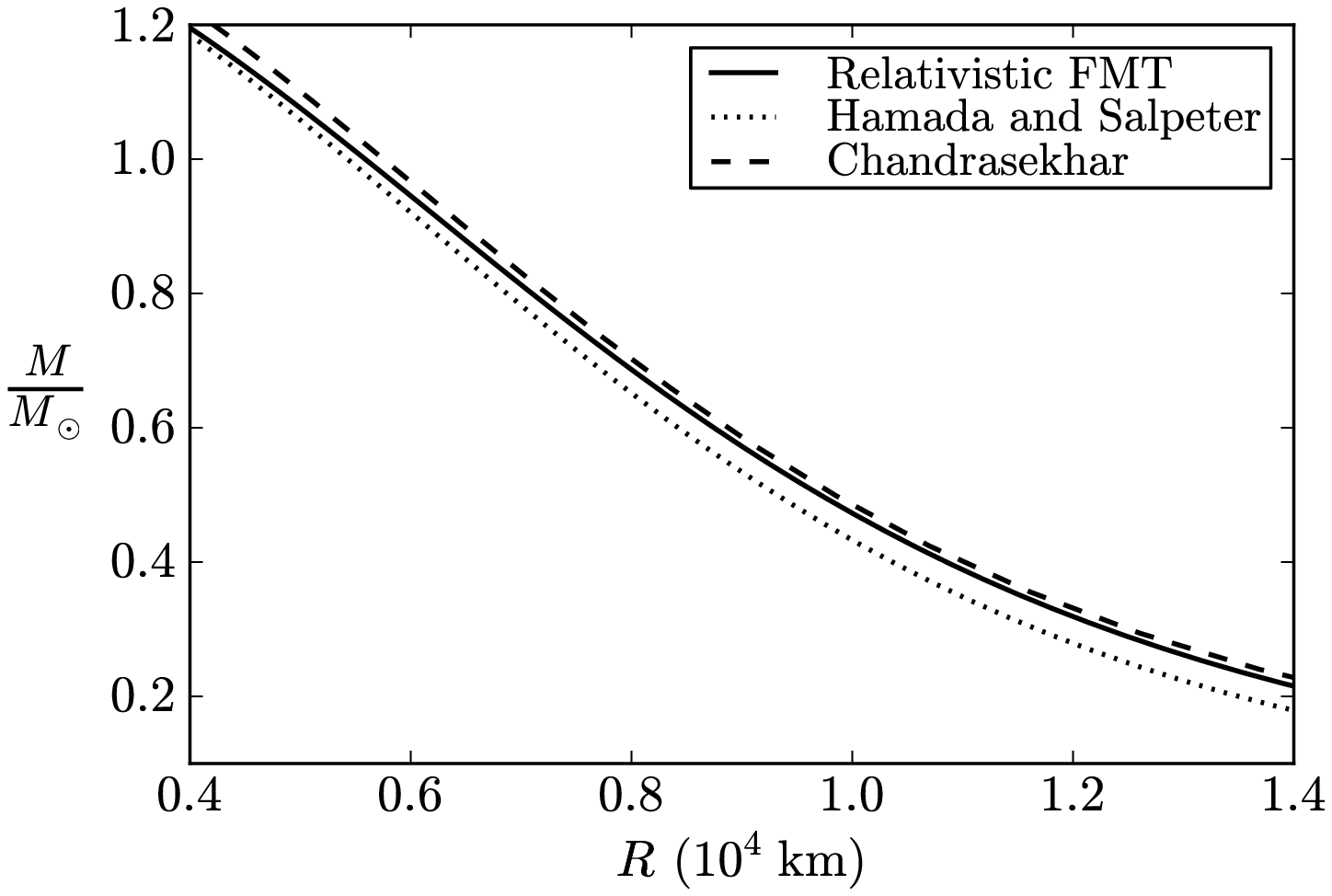} & \includegraphics[width=\columnwidth,clip]{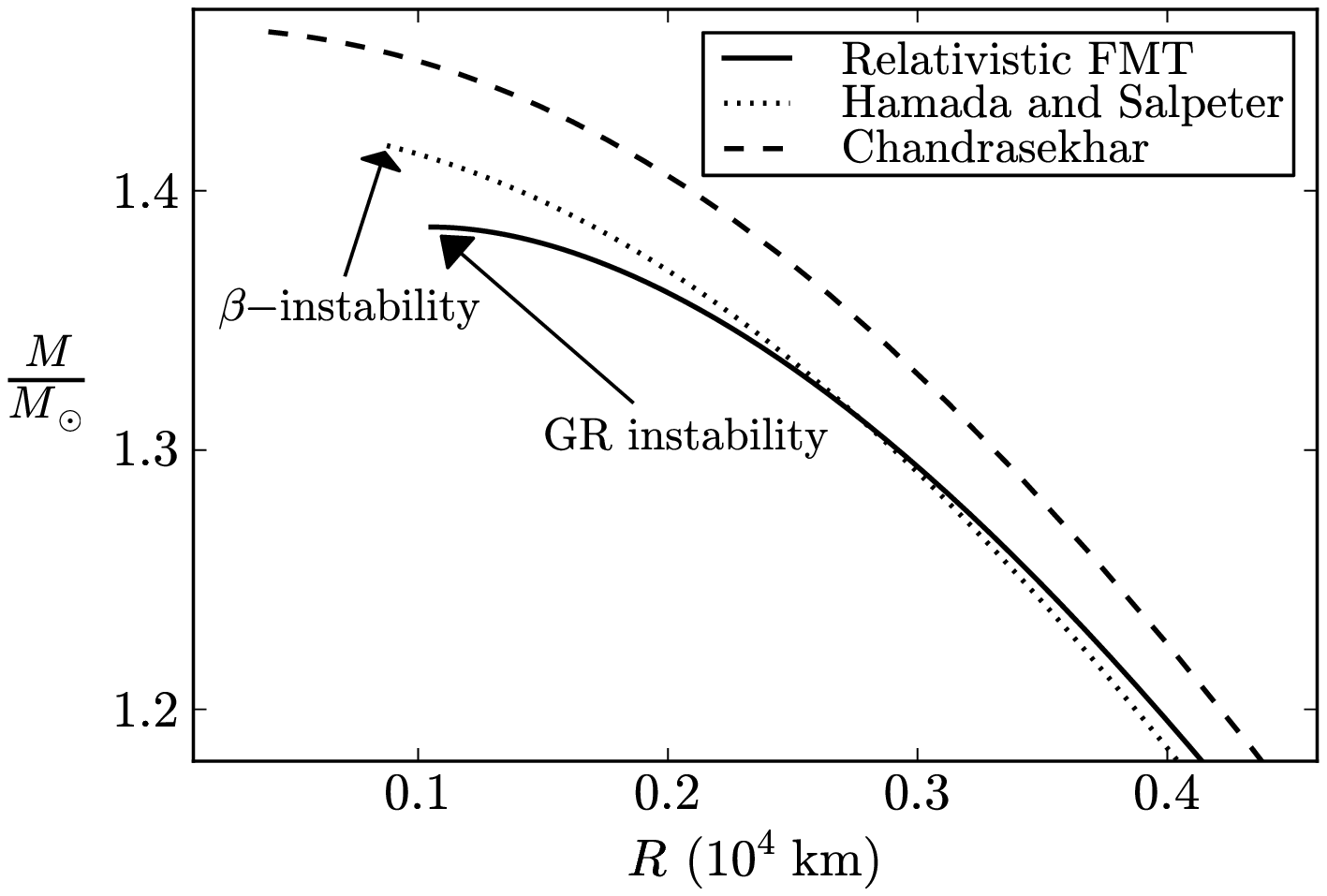}
\end{tabular}
\caption{Mass in solar masses as a function of the radius in units of $10^4$ km for $^{12}$C white dwarfs. The left and right panels show the configurations for the same range of central densities of the corresponding panels of Fig.~\ref{fig:MrhoC}. The solid curve corresponds to the present work, the dotted curves are the Newtonian configurations of Hamada and Salpeter and the dashed curve are the Newtonian configurations of Chandrasekhar.}\label{fig:MRC}
\end{figure*}

\begin{figure*}
\centering
\begin{tabular}{lr}
\includegraphics[width=\columnwidth,clip]{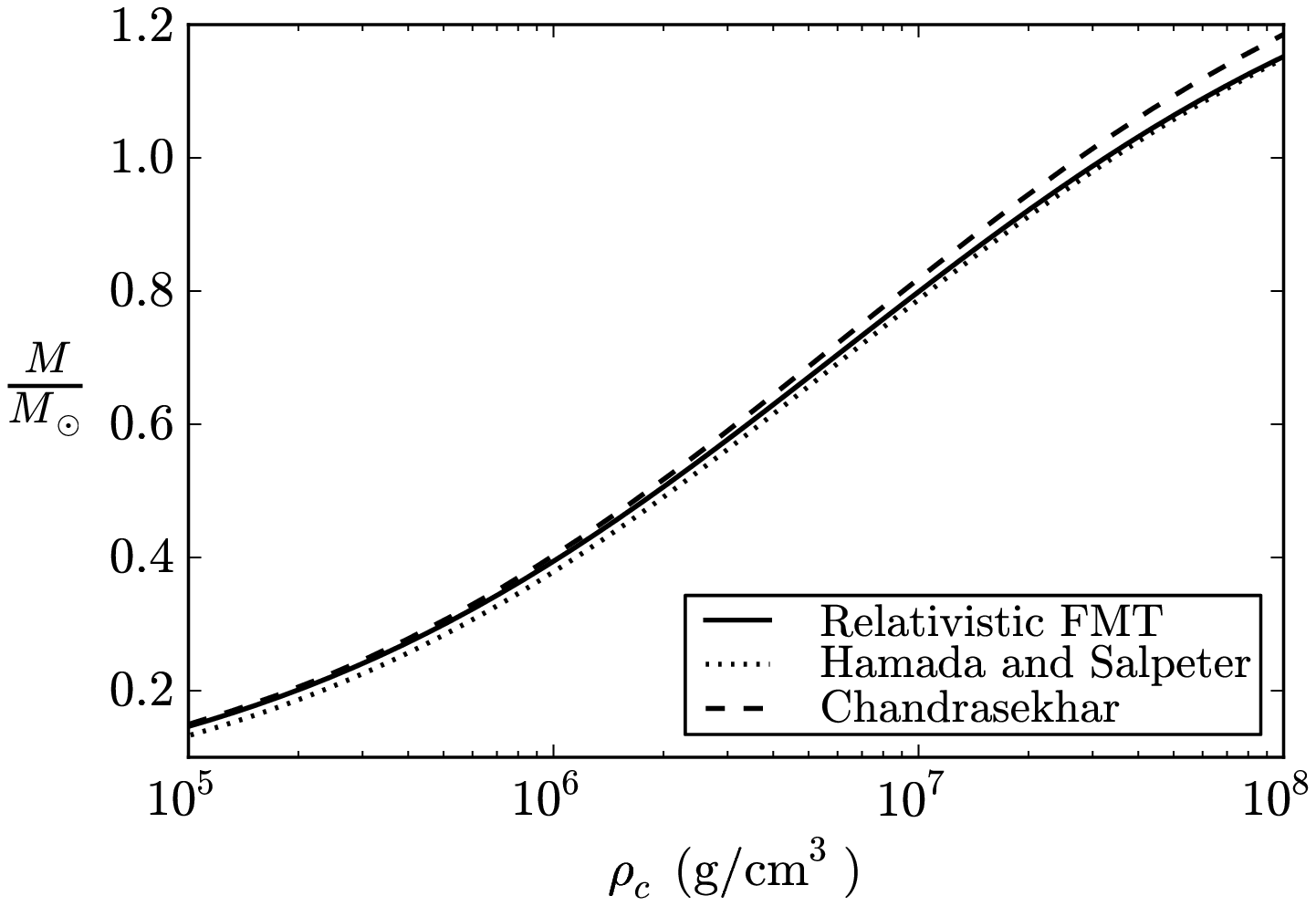} & \includegraphics[width=\columnwidth,clip]{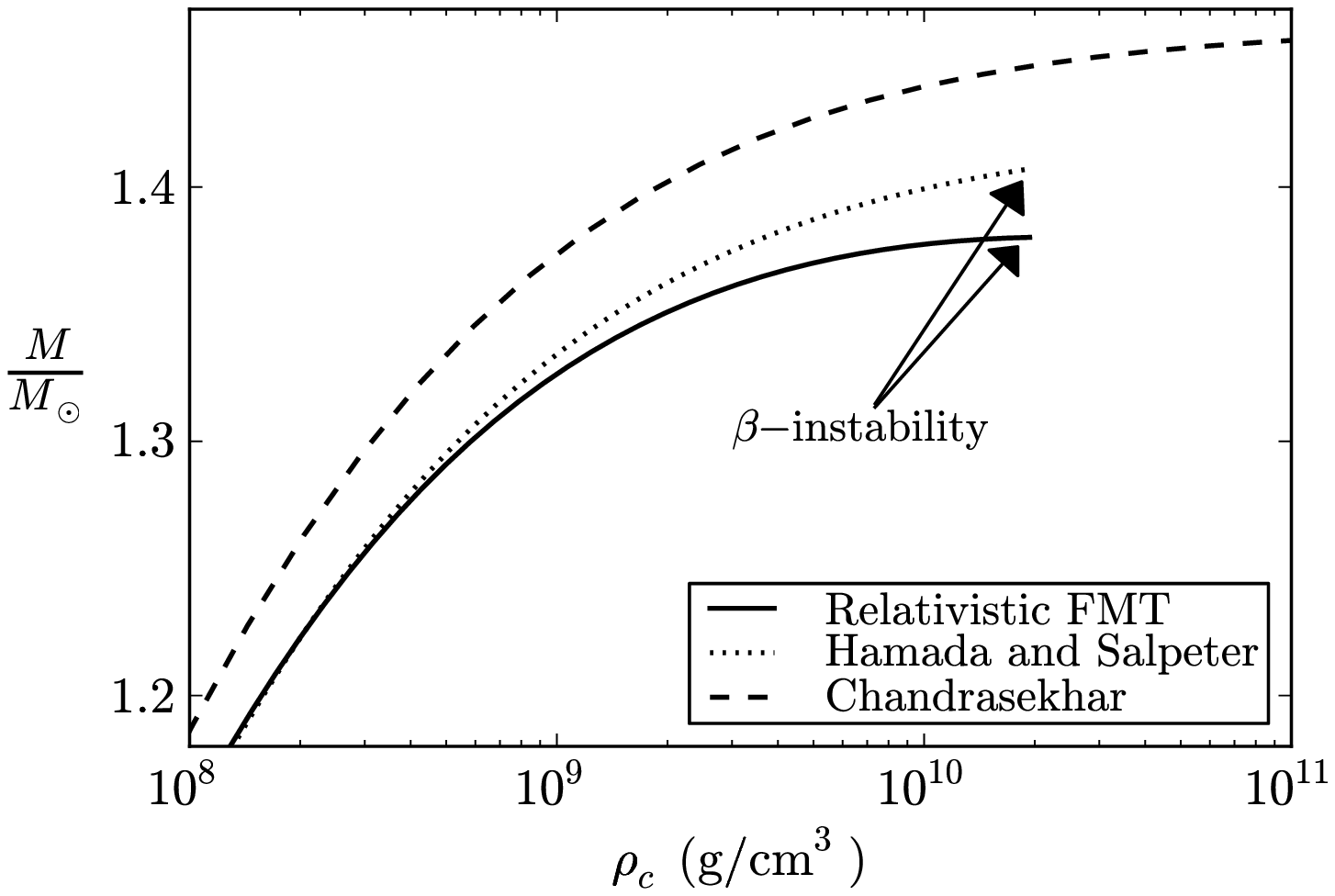}
\end{tabular}
\caption{Mass in solar masses as a function of the central density in the range (left panel) $10^5$--$10^8$ g/cm$^3$ and in the range (right panel) $10^8$--$10^{11}$ g/cm$^3$ for $^{16}$O white dwarfs. The solid curve corresponds to the present work, the dotted curves are the Newtonian configurations of Hamada and Salpeter and the dashed curve are the Newtonian configurations of Chandrasekhar.}\label{fig:MrhoO}
\end{figure*}

\begin{figure*}
\centering
\begin{tabular}{lr}
\includegraphics[width=\columnwidth,clip]{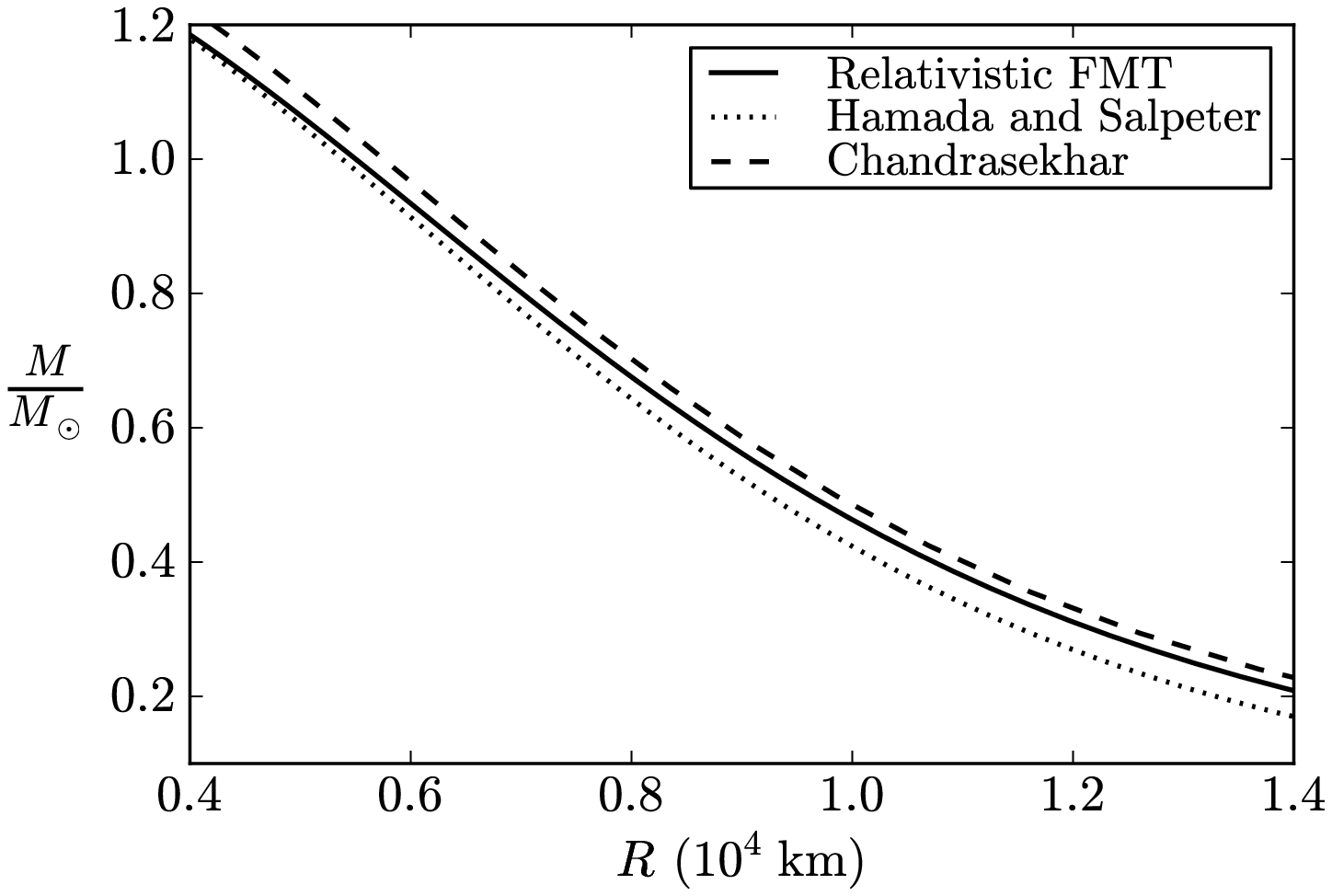} & \includegraphics[width=\columnwidth,clip]{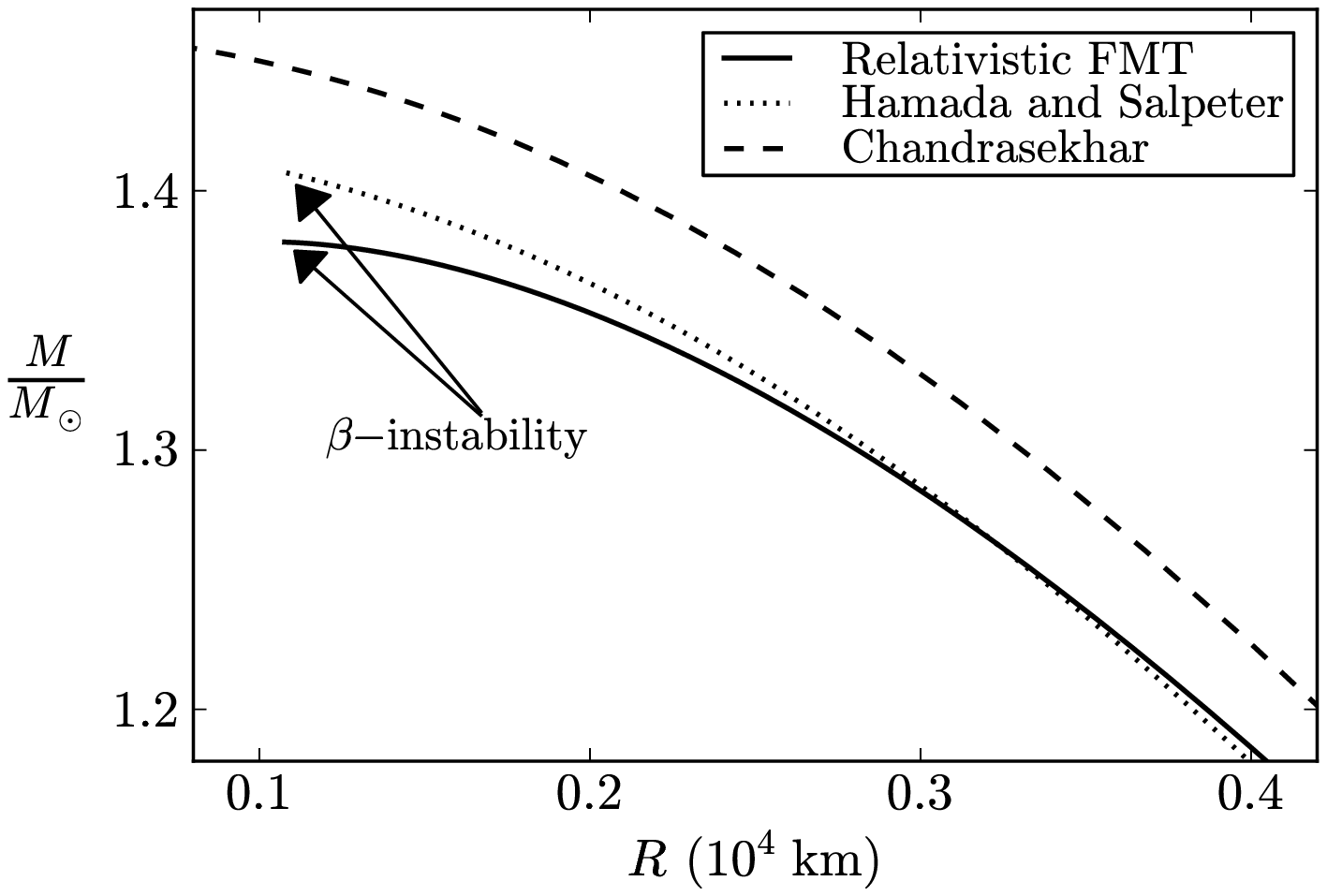}
\end{tabular}
\caption{Mass in solar masses as a function of the radius in units of $10^4$ km for $^{16}$O white dwarfs. The left and right panels show the configurations for the same range of central densities of the corresponding panels of Fig.~\ref{fig:MrhoO}. The solid curve corresponds to the present work, the dotted curves are the Newtonian configurations of Hamada and Salpeter and the dashed curve are the Newtonian configurations of Chandrasekhar.}\label{fig:MRO}
\end{figure*}

\begin{figure*}
\centering
\begin{tabular}{lr}
\includegraphics[width=\columnwidth,clip]{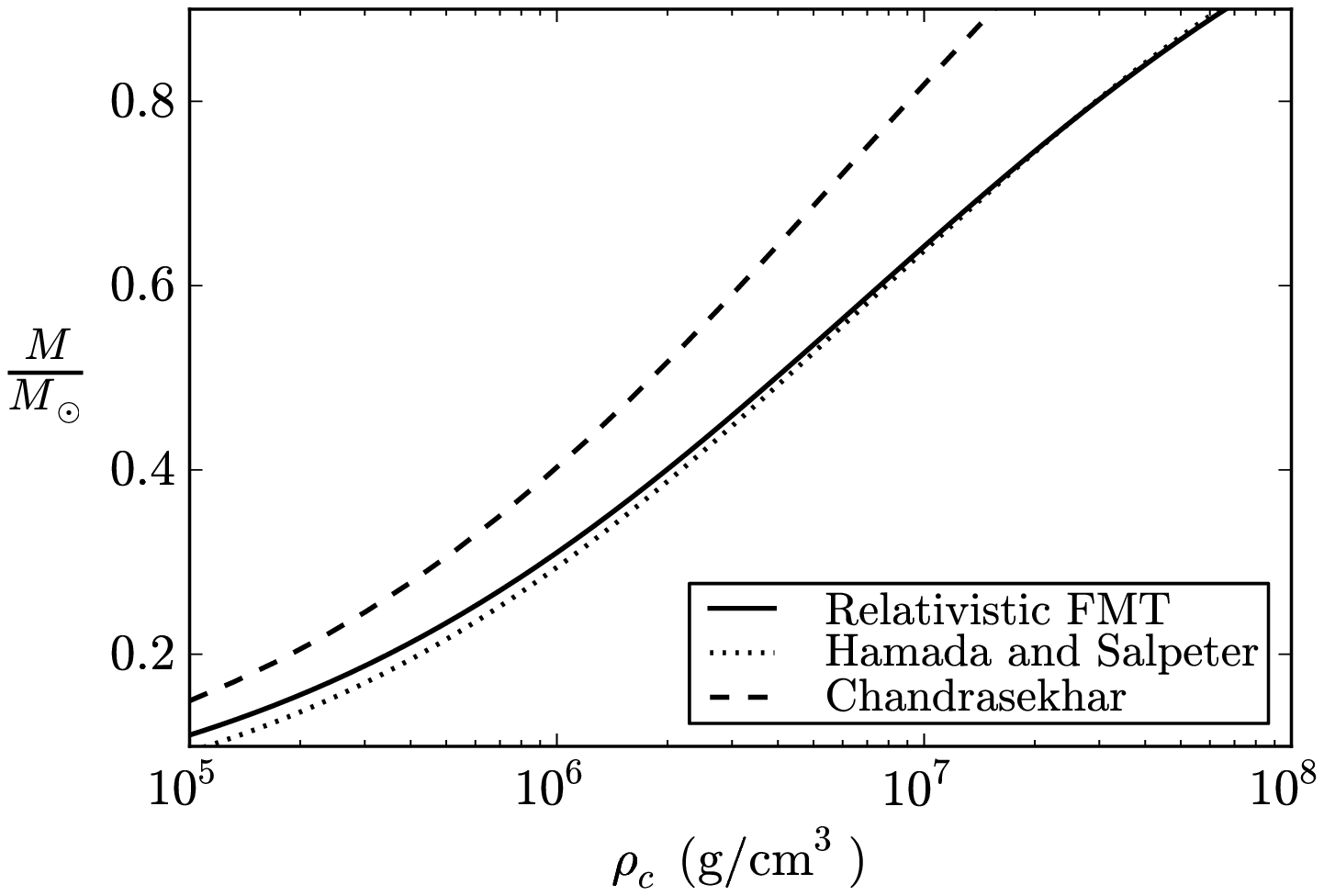} & \includegraphics[width=\columnwidth,clip]{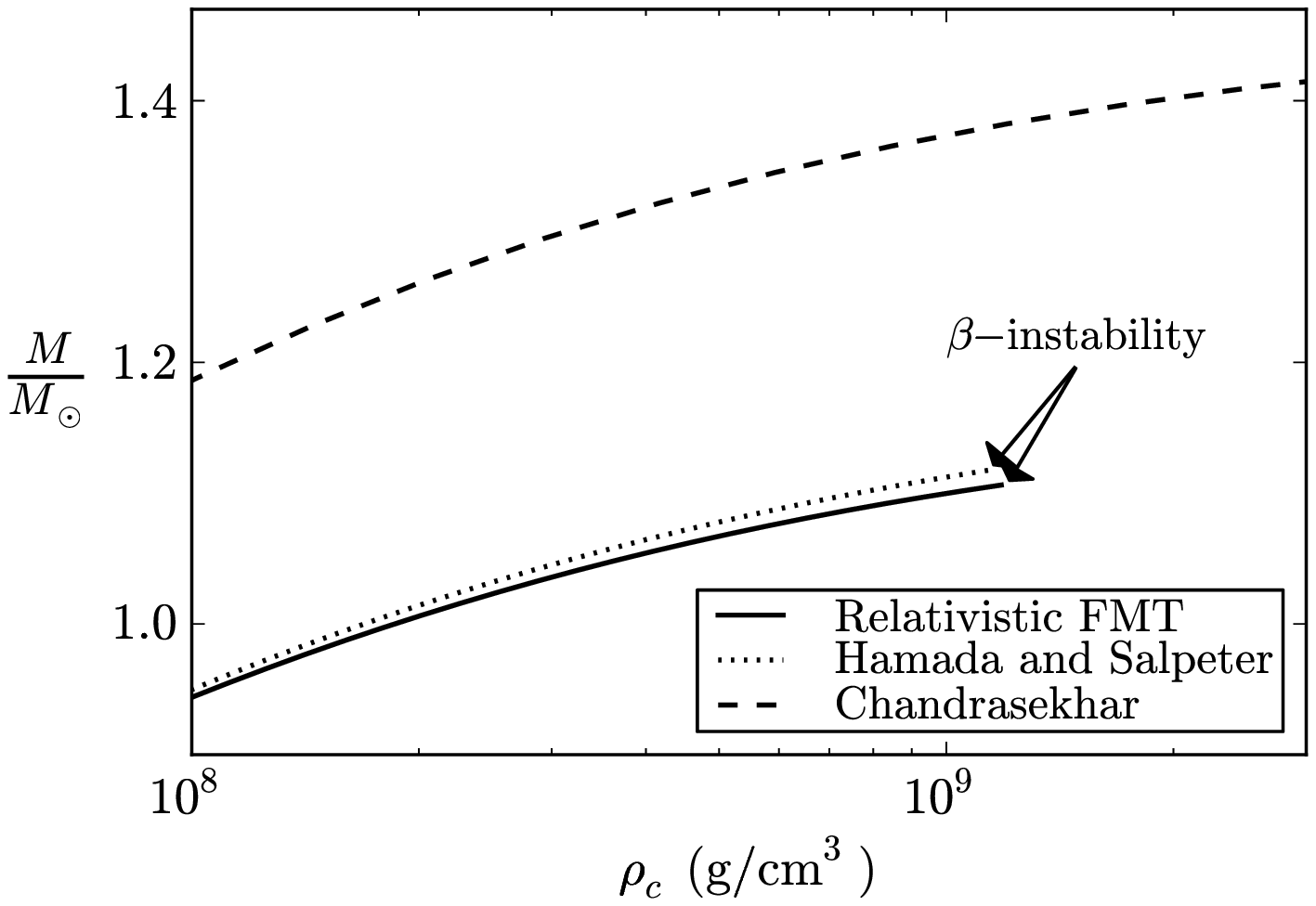}
\end{tabular}
\caption{Mass in solar masses as a function of the central density in the range (left panel) $10^5$--$10^8$ g/cm$^3$ and in the range (right panel) $10^8$--$3\times 10^{9}$ g/cm$^3$ for $^{56}$Fe white dwarfs. The solid curve corresponds to the present work, the dotted curves are the Newtonian configurations of Hamada and Salpeter and the dashed curve are the Newtonian configurations of Chandrasekhar.}\label{fig:MrhoFe}
\end{figure*}

\begin{figure*}
\centering
\begin{tabular}{lr}
\includegraphics[width=\columnwidth,clip]{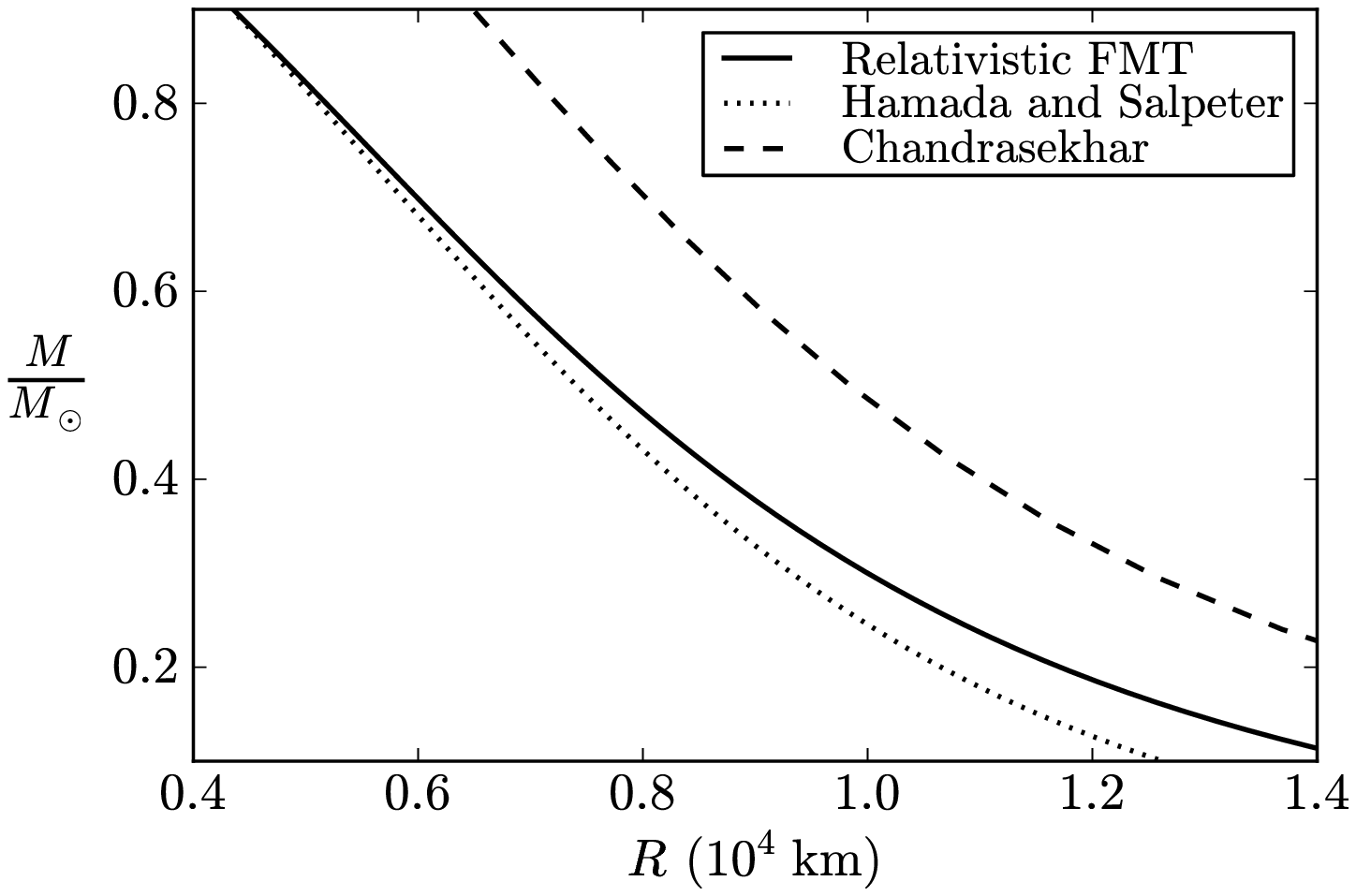} & \includegraphics[width=\columnwidth,clip]{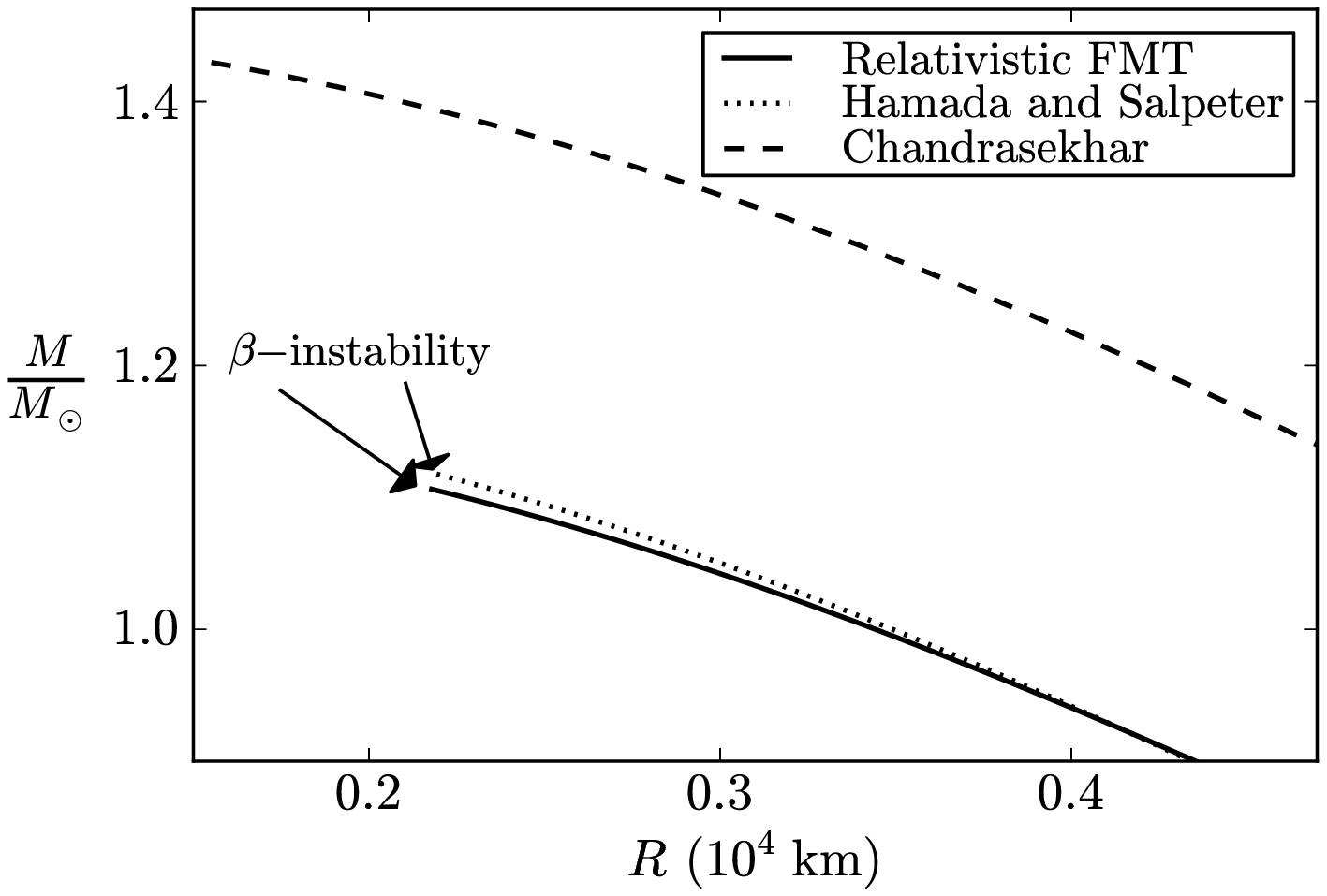}
\end{tabular}
\caption{Mass in solar masses as a function of the radius in units of $10^4$ km for $^{56}$Fe white dwarfs. The left and right panels show the configurations for the same range of central densities of the corresponding panels of Fig.~\ref{fig:MrhoFe}. The solid curve corresponds to the present work, the dotted curves are the Newtonian configurations of Hamada and Salpeter and the dashed curve are the Newtonian configurations of Chandrasekhar.}\label{fig:MRFe}
\end{figure*}

In Figs.~\ref{fig:MrhoHe}--\ref{fig:MRFe} we have plotted the mass-central density relation and the mass-radius relation of general relativistic $^{4}$He, $^{12}$C, $^{16}$O and $^{56}$Fe white dwarfs. In particular, we show the results for the Newtonian white dwarfs of Hamada and Salpeter \cite{hamada61}, for the Newtonian white dwarfs of Chandrasekhar \cite{chandrasekhar31} and the general relativistic configurations obtained in this work based on the relativistic Feynman-Metropolis-Teller equation of state \cite{2011PhRvC..83d5805R}.

Since our approach takes into account self-consistently both $\beta$-decay equilibrium and general relativity, we can determine if the critical mass is reached due either to inverse $\beta$-decay instability or to the general relativistic instability. 

A comparison of the numerical value of the critical mass as given by Stoner \cite{stoner29}, Eq.~(\ref{eq:maxmassStoner}), by Chandrasekhar \cite{chandrasekhar31} and Landau \cite{landau32}, Eq.~(\ref{eq:maxmass}), by Hamada and Salpeter \cite{hamada61} and, by the treatment presented here can be found in Table \ref{tab:mcrit}.

\begin{table}[floatfix]
\begin{center}
\begin{ruledtabular}
\begin{tabular}{c c c c c}
& $\rho^{\rm H \& S}_{\rm crit}$ & $M^{\rm H \& S}_{\rm crit}/M_\odot$ & $\rho^{\rm FMTrel}_{\rm crit}$ & $M^{\rm FMTrel}_{\rm crit}/M_\odot$\\
\hline $^{4}$He & $1.37\times 10^{11}$ & 1.44064 & $1.56\times 10^{10}$ & 1.40906 \\
$^{12}$C & $3.88\times 10^{10}$ & 1.41745 & $2.12\times 10^{10}$ & 1.38603 \\
$^{16}$O & $1.89\times 10^{10}$ & 1.40696 & $1.94\times 10^{10}$ & 1.38024 \\
$^{56}$Fe & $1.14\times 10^9$ & 1.11765 & $1.18\times 10^9$ & 1.10618
\end{tabular}
\end{ruledtabular}
\caption{Critical density and corresponding critical mass for the onset of gravitational collapse of the Newtonian $^4$He, $^{12}$C, $^{16}$O and $^{56}$Fe white dwarfs of Hamada \cite{hamada61}, based on the Salpeter equation of state \cite{salpeter61}, and of the corresponding general relativistic configurations obtained in this work based on the relativistic Feynman-Metropolis-Teller equation of state \cite{2011PhRvC..83d5805R}. Densities are in g/cm$^3$ and masses in solar masses. For the sake of comparison, the critical mass of Stoner (\ref{eq:maxmassStoner}) and of the one of Chandrasekhar-Landau (\ref{eq:maxmass}) are $M^{\rm Stoner}_{\rm crit} \sim 1.72 M_\odot$ and $M^{\rm Ch-L}_{\rm crit}\sim 1.45 M_\odot$, for the average molecular weight $\mu = A_r/Z = 2$.}
\label{tab:mcrit}
\end{center}
\end{table}

From the numerical integrations we have obtained:

\begin{enumerate}

\item $^4$He and $^{12}$C white dwarfs satisfy $\rho^{\rm GR}_{\rm crit} < \rho^{\beta}_{\rm crit}$ (see Figs.~\ref{fig:MrhoHe}--\ref{fig:MRC} and Tables \ref{tab:betadecay} and \ref{tab:mcrit}), so they are unstable with respect to general relativistic effects. The critical density of $^{12}$C white dwarfs is $\sim 2.12\times 10^{10}$ g/cm$^3$, to be compared with the value $2.65\times 10^{10}$ g/cm$^3$ obtained from calculations based on general relativistic corrections to the theory of polytropes (see e.g.~\cite{shapirobook}). 

\item White dwarfs composed of heavier material than $^{12}$C, e.g.~$^{16}$O and $^{56}$Fe are unstable due to inverse $\beta$-decay of the nuclei (see Figs.~\ref{fig:MrhoO}--\ref{fig:MRFe} and Tables \ref{tab:betadecay} and \ref{tab:mcrit}). It is worth to notice that the correct evaluation of general relativistic effects and of the combined contribution of the electrons to the energy-density of the system introduce, for $^{12}$C white dwarfs, a critical mass not due to the inverse beta decay. When the contribution of the electrons to the energy-density is neglected (e.g.~Chandrasekhar \cite{chandrasekhar31} and Hamada and Salpeter \cite{hamada61}, see Eq.~(\ref{eq:Ech2})) the critical density for Carbon white dwarfs is determined by inverse beta decay irrespective of the effects of general relativity.

\item It can be seen from Figs.~\ref{fig:MrhoHe}--\ref{fig:MRFe} that the drastic decrease of the Salpeter pressure at low densities (see \cite{salpeter61,2011PhRvC..83d5805R} and Table \ref{tab:eos} for details) produces an underestimate of the mass and the radius of low density (low mass) white dwarfs.

\item The Coulomb effects are much more pronounced in the case of white dwarfs with heavy nuclear compositions e.g.~$^{56}$Fe (see Figs.~\ref{fig:MrhoFe} and \ref{fig:MRFe}).

\end{enumerate}

%%%%%%%%%%%%%%%%%%%%%%%%%%%%%%%%%%%%%%%%%%%%%%%%%%%%%%%%%%%%%%
%%%%%%%%%%%%%%%%%%%%%%%%%%%%%%%%%%%%%%%%%%%%%%%%%%%%%%%%%%%%%%
\section{Conclusions}\label{sec:5}
%%%%%%%%%%%%%%%%%%%%%%%%%%%%%%%%%%%%%%%%%%%%%%%%%%%%%%%%%%%%%%
%%%%%%%%%%%%%%%%%%%%%%%%%%%%%%%%%%%%%%%%%%%%%%%%%%%%%%%%%%%%%%

We have addressed the theoretical physics aspects of the white dwarf configurations of equilibrium, quite apart from the astrophysical application.

%In the introduction we have recalled how the study of white dwarfs has often stimulated and taken advantage of crucial progress in theoretical physics and applied mathematics. It is clear that the early considerations of the critical mass of a white dwarf were routed in the concept of quantum statistics and the fermion exclusion principle considered by Fowler \cite{fowler26} and Stoner \cite{stoner24} leading to the critical mass (\ref{eq:maxmassStoner}) \cite{stoner29}. 
%
%The following progress was made by Chandrasekhar \cite{chandrasekhar31} and Landau \cite{landau32} adopting the work by Emden \cite{emdenbook} on the solution of the nonlinear Lane-Emden polytropic differential equations. They obtained the critical mass given by Eq.~(\ref{eq:maxmass}).
%
%It was Salpeter \cite{salpeter61} and later Hamada and Salpeter \cite{hamada61} who brought to full fruition the additional conceptual theoretical physics concept of Wigner-Seitz cells. Salpeter adopted the Wigner-Seitz cell for the description of white dwarf material and studied the perturbation to the uniform electron distribution, given by the Coulomb interactions and their special relativity corrections. The value of the critical mass, although obtained only through numerical integration, can be expressed approximately by Eq.~(\ref{eq:maxmassHS}).

%Still many inconsistencies existed in the theoretical model. 

The recently accomplished description of a compressed atom within the global approach of the relativistic Feynman, Metropolis and Teller \cite{2011PhRvC..83d5805R} has been here solved within the Wigner-Seitz cell and applied to the construction of white dwarfs in the framework of general relativity. From a theoretical physics point of view, this is the first unified approach of white dwarfs taking into account consistently the gravitational, the weak, the strong and the electromagnetic interactions, and it answers open theoretical physics issues in this matter. No analytic formula for the critical mass of white dwarfs can be derived and, on the contrary, the critical mass can obtained only through the numerical integration of the general relativistic equations of equilibrium together with the relativistic Feynman-Metropolis-Teller equation of state.

The value of the critical mass and the radius of white dwarfs in our treatment and in the Hamada and Salpeter \cite{hamada61} treatment becomes a function of the composition of the star. Specific examples have been given in the case of white dwarfs composed of $^4$He, $^{12}$C, $^{16}$O and $^{56}$Fe. The results of Chandrasekhar, of Hamada and Salpeter and ours have been compared and contrasted (see Table \ref{tab:mcrit} and Figs.~\ref{fig:MrhoHe}--\ref{fig:MRFe}). 

The critical mass is a decreasing function of $Z$ and Coulomb effects are more important for heavy nuclear compositions. The validity of the Salpeter approximate formulas increases also with $Z$, namely for heavy nuclear compositions the numerical values of the masses as well as of the radii of white dwarfs obtained using the Salpeter equation of state are closer to the ones obtained from the full numerical integration of the general relativistic treatment presented here.

Turning now to astrophysics, the critical mass of white dwarfs is today acquiring a renewed interest in view of its central role in the explanation of the supernova phenomena \cite{phillips93,riess98,perlmutter99,riess04}. The central role of the critical mass of white dwarfs as related to supernova was presented by F.~Hoyle and W.~A.~Fowler \cite{hoyle60} explaining the difference between type I and type II Supernova. This field has developed in the intervening years to a topic of high precision research in astrophysics and, very likely, both the relativistic and the Coulomb effects outlined in this article will become topic of active confrontation between theory and observation. For instance, the underestimate of the mass and the radius of low density white dwarfs within the Hamada and Salpeter treatment \cite{hamada61} (see Figs.~\ref{fig:MrhoHe}--\ref{fig:MRFe}) leads to the possibility of a direct confrontation with observations in the case of low mass white dwarfs e.g. the companion of the Pulsar J1141-6545 \cite{kramerprivate}.

%Paradoxically, the concept of critical mass was not pursued by Chandrasekhar in order to explain the supernova phenomena, on the contrary, Chandrasekhar purported on the role of the critical mass in discriminating star masses leading to the formation of a white dwarf versus stars never reaching a configuration of equilibrium due to radiation pressure \cite{arnettprivate,giacconi78} \footnote{Chandrasekhar, in an interview with S.~Weart \cite{weart77},  recognized ``... at first I didn't understand what this limit meant and I didn't know how it would end, and how it related to the 3/2 low mass polytropes ...}.

We have finally obtained a general formula in Eq.~(\ref{eq:conslaw2}) as a ``first integral'' of the general relativistic equations of equilibrium. This formula relates the chemical potential of the Wigner-Seitz cells, duly obtained from the relativistic Feynman-Metropolis-Teller model \cite{2011PhRvC..83d5805R} taking into account weak, nuclear and electromagnetic interactions, to the general relativistic gravitational potential at each point of the configuration. Besides its esthetic value, this is an important tool to examine the radial dependence of the white dwarf properties and it can be also applied to the crust of a neutron star as it approaches to the physical important regime of neutron star cores. 

The formalism we have introduced allows in principle to evaluate subtle effects of a nuclear density distribution as a function of the radius and of the Fermi energy of the electrons and of the varying depth of the general relativistic gravitational potential. The theoretical base presented in this article establishes also the correct framework for the formulation of the more general case when finite temperatures and magnetic fields are present. This treatment naturally opens the way to a more precise description of the crust of neutron stars, which will certainly become an active topic of research in view of the recent results by S. Goriely et al. \cite{2011A&A...531A..78G,2011arXiv1107.0899G} and by J.~M.~Pearson et al. \cite{2011PhRvC..83f5810P} on the importance of the Coulomb effects in the r-process nucleosynthesis of the crust material during its post-ejection evolution in the process of gravitational collapse and/or in the merging of neutron star binaries.

%We have explicitly discussed many theoretical issues open for years on white dwarfs, including the legitimately posed by A.~Eddington \cite{eddington35}.

\begin{acknowledgments}
We would like to thank Professor D.~Arnett for discussions.
\end{acknowledgments}

%%%%%%%%%%%%%%%%%%%%%%%%%%%%%%%%%%%%%%%%%%%%%%%%%%%%%%%%%%%%%%
%%											 References                         %%
%%%%%%%%%%%%%%%%%%%%%%%%%%%%%%%%%%%%%%%%%%%%%%%%%%%%%%%%%%%%%%

\bibliographystyle{apsrev}
\bibliography{biblio}

\begin{thebibliography}{55}
\expandafter\ifx\csname natexlab\endcsname\relax\def\natexlab#1{#1}\fi
\expandafter\ifx\csname bibnamefont\endcsname\relax
  \def\bibnamefont#1{#1}\fi
\expandafter\ifx\csname bibfnamefont\endcsname\relax
  \def\bibfnamefont#1{#1}\fi
\expandafter\ifx\csname citenamefont\endcsname\relax
  \def\citenamefont#1{#1}\fi
\expandafter\ifx\csname url\endcsname\relax
  \def\url#1{\texttt{#1}}\fi
\expandafter\ifx\csname urlprefix\endcsname\relax\def\urlprefix{URL }\fi
\providecommand{\bibinfo}[2]{#2}
\providecommand{\eprint}[2][]{\url{#2}}

\bibitem[{\citenamefont{{Fowler}}(1926)}]{fowler26}
\bibinfo{author}{\bibfnamefont{R.~H.} \bibnamefont{{Fowler}}},
  \bibinfo{journal}{\mnras} \textbf{\bibinfo{volume}{87}}, \bibinfo{pages}{114}
  (\bibinfo{year}{1926}).

\bibitem[{\citenamefont{{Stoner}}(1929)}]{stoner29}
\bibinfo{author}{\bibfnamefont{E.~C.} \bibnamefont{{Stoner}}},
  \bibinfo{journal}{Philosophical Magazine (Series 7)}
  \textbf{\bibinfo{volume}{7}}, \bibinfo{pages}{63} (\bibinfo{year}{1929}).

\bibitem[{\citenamefont{{Chandrasekhar}}(1931{\natexlab{a}})}]{chandrasekhar31}
\bibinfo{author}{\bibfnamefont{S.}~\bibnamefont{{Chandrasekhar}}},
  \bibinfo{journal}{\apj} \textbf{\bibinfo{volume}{74}}, \bibinfo{pages}{81}
  (\bibinfo{year}{1931}{\natexlab{a}}).

\bibitem[{\citenamefont{{Milne}}(1930)}]{milne30}
\bibinfo{author}{\bibfnamefont{E.~A.} \bibnamefont{{Milne}}},
  \bibinfo{journal}{\mnras} \textbf{\bibinfo{volume}{91}}, \bibinfo{pages}{4}
  (\bibinfo{year}{1930}).

\bibitem[{\citenamefont{{Emden}}(1907)}]{emdenbook}
\bibinfo{author}{\bibfnamefont{R.}~\bibnamefont{{Emden}}},
  \emph{\bibinfo{title}{{Gaskugeln Anwendungen der Mechanischen W\"armetheorie
  auf Kosmologische und Meteorologische Probleme}}}
  (\bibinfo{publisher}{Leipzig, Teubner}, \bibinfo{address}{Berlin},
  \bibinfo{year}{1907}).

\bibitem[{\citenamefont{{Landau}}(1932)}]{landau32}
\bibinfo{author}{\bibfnamefont{L.~D.} \bibnamefont{{Landau}}},
  \bibinfo{journal}{Phys. Z. Sowjetunion} \textbf{\bibinfo{volume}{1}},
  \bibinfo{pages}{285} (\bibinfo{year}{1932}).

\bibitem[{\citenamefont{{Wali}}(1982)}]{wali82}
\bibinfo{author}{\bibfnamefont{K.~C.} \bibnamefont{{Wali}}},
  \bibinfo{journal}{Physics Today} \textbf{\bibinfo{volume}{35}},
  \bibinfo{pages}{33} (\bibinfo{year}{1982}).

\bibitem[{\citenamefont{{Boccaletti} and {Ruffini}}(2010)}]{ruffinibook}
\bibinfo{author}{\bibfnamefont{D.}~\bibnamefont{{Boccaletti}}}
  \bibnamefont{and}
  \bibinfo{author}{\bibfnamefont{R.}~\bibnamefont{{Ruffini}}},
  \emph{\bibinfo{title}{{Fermi and Astrophysics}}} (\bibinfo{publisher}{World
  Scientific}, \bibinfo{address}{Singapore}, \bibinfo{year}{2010}).

\bibitem[{\citenamefont{{Salpeter}}(1961)}]{salpeter61}
\bibinfo{author}{\bibfnamefont{E.~E.} \bibnamefont{{Salpeter}}},
  \bibinfo{journal}{\apj} \textbf{\bibinfo{volume}{134}}, \bibinfo{pages}{669}
  (\bibinfo{year}{1961}).

\bibitem[{\citenamefont{{Frenkel}}(1928)}]{frenkel28}
\bibinfo{author}{\bibfnamefont{Y.~I.} \bibnamefont{{Frenkel}}},
  \bibinfo{journal}{Zeit. fur Phys.} \textbf{\bibinfo{volume}{50}},
  \bibinfo{pages}{234} (\bibinfo{year}{1928}).

\bibitem[{\citenamefont{{Hamada} and {Salpeter}}(1961)}]{hamada61}
\bibinfo{author}{\bibfnamefont{T.}~\bibnamefont{{Hamada}}} \bibnamefont{and}
  \bibinfo{author}{\bibfnamefont{E.~E.} \bibnamefont{{Salpeter}}},
  \bibinfo{journal}{\apj} \textbf{\bibinfo{volume}{134}}, \bibinfo{pages}{683}
  (\bibinfo{year}{1961}).

\bibitem[{\citenamefont{{Olson} and {Bailyn}}(1975)}]{olson75}
\bibinfo{author}{\bibfnamefont{E.}~\bibnamefont{{Olson}}} \bibnamefont{and}
  \bibinfo{author}{\bibfnamefont{M.}~\bibnamefont{{Bailyn}}},
  \bibinfo{journal}{\prd} \textbf{\bibinfo{volume}{12}}, \bibinfo{pages}{3030}
  (\bibinfo{year}{1975}).

\bibitem[{\citenamefont{{Olson} and {Bailyn}}(1976)}]{olson76}
\bibinfo{author}{\bibfnamefont{E.}~\bibnamefont{{Olson}}} \bibnamefont{and}
  \bibinfo{author}{\bibfnamefont{M.}~\bibnamefont{{Bailyn}}},
  \bibinfo{journal}{\prd} \textbf{\bibinfo{volume}{13}}, \bibinfo{pages}{2204}
  (\bibinfo{year}{1976}).

\bibitem[{\citenamefont{{Rotondo}
  et~al.}(2011{\natexlab{a}})\citenamefont{{Rotondo}, {Rueda}, {Ruffini}, and
  {Xue}}}]{PLB2011}
\bibinfo{author}{\bibfnamefont{M.}~\bibnamefont{{Rotondo}}},
  \bibinfo{author}{\bibfnamefont{J.~A.} \bibnamefont{{Rueda}}},
  \bibinfo{author}{\bibfnamefont{R.}~\bibnamefont{{Ruffini}}},
  \bibnamefont{and} \bibinfo{author}{\bibfnamefont{S.}~\bibnamefont{{Xue}}},
  \bibinfo{journal}{\plb} \textbf{\bibinfo{volume}{701}}, \bibinfo{pages}{667}
  (\bibinfo{year}{2011}{\natexlab{a}}).

\bibitem[{\citenamefont{{Ruffini}}(2000)}]{gursky2000}
\bibinfo{author}{\bibfnamefont{R.}~\bibnamefont{{Ruffini}}}, in
  \emph{\bibinfo{booktitle}{Exploring the universe: a Festschrift in honor of
  Riccardo Giacconi}}, edited by \bibinfo{editor}{\bibnamefont{{H.~Gursky,
  R.~Ruffini, \& L.~Stella}}} (\bibinfo{year}{2000}).

\bibitem[{\citenamefont{{Bertone} and {Ruffini}}(2000)}]{bertone2000}
\bibinfo{author}{\bibfnamefont{G.}~\bibnamefont{{Bertone}}} \bibnamefont{and}
  \bibinfo{author}{\bibfnamefont{R.}~\bibnamefont{{Ruffini}}},
  \bibinfo{journal}{Nuovo Cimento B Serie} \textbf{\bibinfo{volume}{115}},
  \bibinfo{pages}{935} (\bibinfo{year}{2000}).

\bibitem[{\citenamefont{{Rotondo}
  et~al.}(2011{\natexlab{b}})\citenamefont{{Rotondo}, {Rueda}, {Ruffini}, and
  {Xue}}}]{2011PhRvC..83d5805R}
\bibinfo{author}{\bibfnamefont{M.}~\bibnamefont{{Rotondo}}},
  \bibinfo{author}{\bibfnamefont{J.~A.} \bibnamefont{{Rueda}}},
  \bibinfo{author}{\bibfnamefont{R.}~\bibnamefont{{Ruffini}}},
  \bibnamefont{and} \bibinfo{author}{\bibfnamefont{S.-S.} \bibnamefont{{Xue}}},
  \bibinfo{journal}{\prc} \textbf{\bibinfo{volume}{83}},
  \bibinfo{pages}{045805} (\bibinfo{year}{2011}{\natexlab{b}}),
  \eprint{0911.4622}.

\bibitem[{\citenamefont{{Feynman} et~al.}(1949)\citenamefont{{Feynman},
  {Metropolis}, and {Teller}}}]{feynman49}
\bibinfo{author}{\bibfnamefont{R.~P.} \bibnamefont{{Feynman}}},
  \bibinfo{author}{\bibfnamefont{N.}~\bibnamefont{{Metropolis}}},
  \bibnamefont{and} \bibinfo{author}{\bibfnamefont{E.}~\bibnamefont{{Teller}}},
  \bibinfo{journal}{\pr} \textbf{\bibinfo{volume}{75}}, \bibinfo{pages}{1561}
  (\bibinfo{year}{1949}).

\bibitem[{\citenamefont{{Baym} et~al.}(1971)\citenamefont{{Baym}, {Pethick},
  and {Sutherland}}}]{baym71a}
\bibinfo{author}{\bibfnamefont{G.}~\bibnamefont{{Baym}}},
  \bibinfo{author}{\bibfnamefont{C.}~\bibnamefont{{Pethick}}},
  \bibnamefont{and}
  \bibinfo{author}{\bibfnamefont{P.}~\bibnamefont{{Sutherland}}},
  \bibinfo{journal}{\apj} \textbf{\bibinfo{volume}{170}}, \bibinfo{pages}{299}
  (\bibinfo{year}{1971}).

\bibitem[{\citenamefont{{Landau} and {Lifshitz}}(1980)}]{landaubook}
\bibinfo{author}{\bibfnamefont{L.~D.} \bibnamefont{{Landau}}} \bibnamefont{and}
  \bibinfo{author}{\bibfnamefont{E.~M.} \bibnamefont{{Lifshitz}}},
  \emph{\bibinfo{title}{{Statistical physics. Part1}}}
  (\bibinfo{publisher}{Pergamon Press}, \bibinfo{address}{Oxford},
  \bibinfo{year}{1980}).

\bibitem[{\citenamefont{{Ferreirinho} et~al.}(1980)\citenamefont{{Ferreirinho},
  {Ruffini}, and {Stella}}}]{ferreirinho80}
\bibinfo{author}{\bibfnamefont{J.}~\bibnamefont{{Ferreirinho}}},
  \bibinfo{author}{\bibfnamefont{R.}~\bibnamefont{{Ruffini}}},
  \bibnamefont{and} \bibinfo{author}{\bibfnamefont{L.}~\bibnamefont{{Stella}}},
  \bibinfo{journal}{\plb} \textbf{\bibinfo{volume}{91}}, \bibinfo{pages}{314}
  (\bibinfo{year}{1980}).

\bibitem[{\citenamefont{{Ruffini} and {Stella}}(1981)}]{ruffini81}
\bibinfo{author}{\bibfnamefont{R.}~\bibnamefont{{Ruffini}}} \bibnamefont{and}
  \bibinfo{author}{\bibfnamefont{L.}~\bibnamefont{{Stella}}},
  \bibinfo{journal}{\plb} \textbf{\bibinfo{volume}{102}}, \bibinfo{pages}{442}
  (\bibinfo{year}{1981}).

\bibitem[{\citenamefont{{Dirac}}(1930)}]{dirac30}
\bibinfo{author}{\bibfnamefont{P.~A.~M.} \bibnamefont{{Dirac}}},
  \bibinfo{journal}{Proc.\ Cambridge Phil.\ Soc.}
  \textbf{\bibinfo{volume}{26}}, \bibinfo{pages}{376} (\bibinfo{year}{1930}).

\bibitem[{\citenamefont{{Migdal} et~al.}(1977)\citenamefont{{Migdal}, {Popov},
  and {Voskresenski{\v i}}}}]{migdal77}
\bibinfo{author}{\bibfnamefont{A.~B.} \bibnamefont{{Migdal}}},
  \bibinfo{author}{\bibfnamefont{V.~S.} \bibnamefont{{Popov}}},
  \bibnamefont{and} \bibinfo{author}{\bibfnamefont{D.~N.}
  \bibnamefont{{Voskresenski{\v i}}}}, \bibinfo{journal}{Soviet Journal of
  Experimental and Theoretical Physics} \textbf{\bibinfo{volume}{45}},
  \bibinfo{pages}{436} (\bibinfo{year}{1977}).

\bibitem[{\citenamefont{{Slater} and {Krutter}}(1935)}]{slater35}
\bibinfo{author}{\bibfnamefont{J.~C.} \bibnamefont{{Slater}}} \bibnamefont{and}
  \bibinfo{author}{\bibfnamefont{H.~M.} \bibnamefont{{Krutter}}},
  \bibinfo{journal}{Phys. Rev.} \textbf{\bibinfo{volume}{47}},
  \bibinfo{pages}{559} (\bibinfo{year}{1935}).

\bibitem[{\citenamefont{{Ruffini}}(2008)}]{2008pint.conf..207R}
\bibinfo{author}{\bibfnamefont{R.}~\bibnamefont{{Ruffini}}}, in
  \emph{\bibinfo{booktitle}{Path Integrals - New Trends and Perspectives}}
  (\bibinfo{year}{2008}), pp. \bibinfo{pages}{207--218}.

\bibitem[{\citenamefont{{Popov} et~al.}(2011)\citenamefont{{Popov}, {Rotondo},
  {Ruffini}, and {Xue}}}]{popov10}
\bibinfo{author}{\bibfnamefont{V.~S.} \bibnamefont{{Popov}}},
  \bibinfo{author}{\bibfnamefont{M.}~\bibnamefont{{Rotondo}}},
  \bibinfo{author}{\bibfnamefont{R.}~\bibnamefont{{Ruffini}}},
  \bibnamefont{and} \bibinfo{author}{\bibfnamefont{S.}~\bibnamefont{{Xue}}}, in
  \emph{\bibinfo{booktitle}{Proceedings of the 1st Galileo-Xu Guantqi Meeting
  (2009), Int. J. Mod. Phys. Conf. S. in press}}, edited by
  \bibinfo{editor}{\bibnamefont{{D.~Blair, R.~Ruffini, J.~Yipeng and
  S.-S.~Xue}}} (\bibinfo{year}{2011}), \eprint{arXiv:astro-ph/0903.3727}.

\bibitem[{\citenamefont{{Tolman}}(1939)}]{tolman39}
\bibinfo{author}{\bibfnamefont{R.~C.} \bibnamefont{{Tolman}}},
  \bibinfo{journal}{Physical Review} \textbf{\bibinfo{volume}{55}},
  \bibinfo{pages}{364} (\bibinfo{year}{1939}).

\bibitem[{\citenamefont{{Oppenheimer} and {Volkoff}}(1939)}]{oppenheimer39}
\bibinfo{author}{\bibfnamefont{J.~R.} \bibnamefont{{Oppenheimer}}}
  \bibnamefont{and} \bibinfo{author}{\bibfnamefont{G.~M.}
  \bibnamefont{{Volkoff}}}, \bibinfo{journal}{\pr}
  \textbf{\bibinfo{volume}{55}}, \bibinfo{pages}{374} (\bibinfo{year}{1939}).

\bibitem[{\citenamefont{{Klein}}(1949)}]{klein49}
\bibinfo{author}{\bibfnamefont{O.}~\bibnamefont{{Klein}}},
  \bibinfo{journal}{Reviews of Modern Physics} \textbf{\bibinfo{volume}{21}},
  \bibinfo{pages}{531} (\bibinfo{year}{1949}).

\bibitem[{\citenamefont{{Tolman}}(1930)}]{1930PhRv...35..904T}
\bibinfo{author}{\bibfnamefont{R.~C.} \bibnamefont{{Tolman}}},
  \bibinfo{journal}{Physical Review} \textbf{\bibinfo{volume}{35}},
  \bibinfo{pages}{904} (\bibinfo{year}{1930}).

\bibitem[{\citenamefont{{Tolman} and {Ehrenfest}}(1930)}]{1930PhRv...36.1791T}
\bibinfo{author}{\bibfnamefont{R.~C.} \bibnamefont{{Tolman}}} \bibnamefont{and}
  \bibinfo{author}{\bibfnamefont{P.}~\bibnamefont{{Ehrenfest}}},
  \bibinfo{journal}{Physical Review} \textbf{\bibinfo{volume}{36}},
  \bibinfo{pages}{1791} (\bibinfo{year}{1930}).

\bibitem[{\citenamefont{{Chandrasekhar}}(1931{\natexlab{b}})}]{chandrasekhar31%
a}
\bibinfo{author}{\bibfnamefont{S.}~\bibnamefont{{Chandrasekhar}}},
  \bibinfo{journal}{\mnras} \textbf{\bibinfo{volume}{91}}, \bibinfo{pages}{456}
  (\bibinfo{year}{1931}{\natexlab{b}}).

\bibitem[{\citenamefont{{Chandrasekhar}}(1935)}]{chandrasekhar35}
\bibinfo{author}{\bibfnamefont{S.}~\bibnamefont{{Chandrasekhar}}},
  \bibinfo{journal}{\mnras} \textbf{\bibinfo{volume}{95}}, \bibinfo{pages}{207}
  (\bibinfo{year}{1935}).

\bibitem[{\citenamefont{{Chandrasekhar}}(1939)}]{chandrasekharbook}
\bibinfo{author}{\bibfnamefont{S.}~\bibnamefont{{Chandrasekhar}}},
  \emph{\bibinfo{title}{{An introduction to the study of stellar structure}}}
  (\bibinfo{year}{1939}).

\bibitem[{\citenamefont{{Eddington}}(1935)}]{eddington35}
\bibinfo{author}{\bibfnamefont{A.~S.} \bibnamefont{{Eddington}},
  \bibfnamefont{Sir}}, \bibinfo{journal}{\mnras} \textbf{\bibinfo{volume}{95}},
  \bibinfo{pages}{194} (\bibinfo{year}{1935}).

\bibitem[{\citenamefont{{Ciufolini} and {Ruffini}}(1983)}]{ciufolini83}
\bibinfo{author}{\bibfnamefont{I.}~\bibnamefont{{Ciufolini}}} \bibnamefont{and}
  \bibinfo{author}{\bibfnamefont{R.}~\bibnamefont{{Ruffini}}},
  \bibinfo{journal}{\apj} \textbf{\bibinfo{volume}{275}}, \bibinfo{pages}{867}
  (\bibinfo{year}{1983}).

\bibitem[{\citenamefont{{Hund}}(1936)}]{hund36}
\bibinfo{author}{\bibfnamefont{F.}~\bibnamefont{{Hund}}},
  \bibinfo{journal}{Erg.\ d.\ exacten Natwis.} \textbf{\bibinfo{volume}{15}},
  \bibinfo{pages}{189} (\bibinfo{year}{1936}).

\bibitem[{\citenamefont{{Landau}}(1938)}]{landau38}
\bibinfo{author}{\bibfnamefont{L.~D.} \bibnamefont{{Landau}}},
  \bibinfo{journal}{Nature} \textbf{\bibinfo{volume}{19}}, \bibinfo{pages}{333}
  (\bibinfo{year}{1938}).

\bibitem[{\citenamefont{{Zel'Dovich}}(1958)}]{zeldovich58}
\bibinfo{author}{\bibfnamefont{I.~B.} \bibnamefont{{Zel'Dovich}}},
  \bibinfo{journal}{Soviet Journal of Experimental and Theoretical Physics}
  \textbf{\bibinfo{volume}{6}}, \bibinfo{pages}{760} (\bibinfo{year}{1958}).

\bibitem[{\citenamefont{{Harrison} et~al.}(1958)\citenamefont{{Harrison},
  {Wakano}, and {Wheeler}}}]{harrison58}
\bibinfo{author}{\bibfnamefont{B.~K.} \bibnamefont{{Harrison}}},
  \bibinfo{author}{\bibfnamefont{M.}~\bibnamefont{{Wakano}}}, \bibnamefont{and}
  \bibinfo{author}{\bibfnamefont{J.~A.} \bibnamefont{{Wheeler}}},
  \bibinfo{journal}{Onzieme Conseil de Physisque de Solvay}
  (\bibinfo{year}{1958}).

\bibitem[{\citenamefont{{Shapiro} and {Teukolsky}}(1983)}]{shapirobook}
\bibinfo{author}{\bibfnamefont{S.~L.} \bibnamefont{{Shapiro}}}
  \bibnamefont{and} \bibinfo{author}{\bibfnamefont{S.~A.}
  \bibnamefont{{Teukolsky}}}, \emph{\bibinfo{title}{{Black holes, white dwarfs,
  and neutron stars: The physics of compact objects}}} (\bibinfo{year}{1983}).

\bibitem[{\citenamefont{{Audi} et~al.}(2003)\citenamefont{{Audi}, {Wapstra},
  and {Thibault}}}]{audi03}
\bibinfo{author}{\bibfnamefont{G.}~\bibnamefont{{Audi}}},
  \bibinfo{author}{\bibfnamefont{A.~H.} \bibnamefont{{Wapstra}}},
  \bibnamefont{and}
  \bibinfo{author}{\bibfnamefont{C.}~\bibnamefont{{Thibault}}},
  \bibinfo{journal}{Nuclear Physics A} \textbf{\bibinfo{volume}{729}},
  \bibinfo{pages}{337} (\bibinfo{year}{2003}).

\bibitem[{\citenamefont{{Wapstra} and {Bos}}(1977)}]{1977ADNDT..19..175W}
\bibinfo{author}{\bibfnamefont{A.~H.} \bibnamefont{{Wapstra}}}
  \bibnamefont{and} \bibinfo{author}{\bibfnamefont{K.}~\bibnamefont{{Bos}}},
  \bibinfo{journal}{Atomic Data and Nuclear Data Tables}
  \textbf{\bibinfo{volume}{19}}, \bibinfo{pages}{175} (\bibinfo{year}{1977}).

\bibitem[{\citenamefont{{Phillips}}(1993)}]{phillips93}
\bibinfo{author}{\bibfnamefont{M.~M.} \bibnamefont{{Phillips}}},
  \bibinfo{journal}{Astrophys. J. Lett.} \textbf{\bibinfo{volume}{413}},
  \bibinfo{pages}{L105} (\bibinfo{year}{1993}).

\bibitem[{\citenamefont{{Riess} et~al.}(1998)\citenamefont{{Riess},
  {Filippenko}, {Challis}, {Clocchiatti}, {Diercks}, {Garnavich}, {Gilliland},
  {Hogan}, {Jha}, {Kirshner} et~al.}}]{riess98}
\bibinfo{author}{\bibfnamefont{A.~G.} \bibnamefont{{Riess}}},
  \bibinfo{author}{\bibfnamefont{A.~V.} \bibnamefont{{Filippenko}}},
  \bibinfo{author}{\bibfnamefont{P.}~\bibnamefont{{Challis}}},
  \bibinfo{author}{\bibfnamefont{A.}~\bibnamefont{{Clocchiatti}}},
  \bibinfo{author}{\bibfnamefont{A.}~\bibnamefont{{Diercks}}},
  \bibinfo{author}{\bibfnamefont{P.~M.} \bibnamefont{{Garnavich}}},
  \bibinfo{author}{\bibfnamefont{R.~L.} \bibnamefont{{Gilliland}}},
  \bibinfo{author}{\bibfnamefont{C.~J.} \bibnamefont{{Hogan}}},
  \bibinfo{author}{\bibfnamefont{S.}~\bibnamefont{{Jha}}},
  \bibinfo{author}{\bibfnamefont{R.~P.} \bibnamefont{{Kirshner}}},
  \bibnamefont{et~al.}, \bibinfo{journal}{Astronomical Journal}
  \textbf{\bibinfo{volume}{116}}, \bibinfo{pages}{1009} (\bibinfo{year}{1998}),
  \eprint{arXiv:astro-ph/9805201}.

\bibitem[{\citenamefont{{Perlmutter} et~al.}(1999)\citenamefont{{Perlmutter},
  {Aldering}, {Goldhaber}, {Knop}, {Nugent}, {Castro}, {Deustua}, {Fabbro},
  {Goobar}, {Groom} et~al.}}]{perlmutter99}
\bibinfo{author}{\bibfnamefont{S.}~\bibnamefont{{Perlmutter}}},
  \bibinfo{author}{\bibfnamefont{G.}~\bibnamefont{{Aldering}}},
  \bibinfo{author}{\bibfnamefont{G.}~\bibnamefont{{Goldhaber}}},
  \bibinfo{author}{\bibfnamefont{R.~A.} \bibnamefont{{Knop}}},
  \bibinfo{author}{\bibfnamefont{P.}~\bibnamefont{{Nugent}}},
  \bibinfo{author}{\bibfnamefont{P.~G.} \bibnamefont{{Castro}}},
  \bibinfo{author}{\bibfnamefont{S.}~\bibnamefont{{Deustua}}},
  \bibinfo{author}{\bibfnamefont{S.}~\bibnamefont{{Fabbro}}},
  \bibinfo{author}{\bibfnamefont{A.}~\bibnamefont{{Goobar}}},
  \bibinfo{author}{\bibfnamefont{D.~E.} \bibnamefont{{Groom}}},
  \bibnamefont{et~al.}, \bibinfo{journal}{\apj} \textbf{\bibinfo{volume}{517}},
  \bibinfo{pages}{565} (\bibinfo{year}{1999}), \eprint{arXiv:astro-ph/9812133}.

\bibitem[{\citenamefont{{Riess} et~al.}(2004)\citenamefont{{Riess}, {Strolger},
  {Tonry}, {Casertano}, {Ferguson}, {Mobasher}, {Challis}, {Filippenko}, {Jha},
  {Li} et~al.}}]{riess04}
\bibinfo{author}{\bibfnamefont{A.~G.} \bibnamefont{{Riess}}},
  \bibinfo{author}{\bibfnamefont{L.}~\bibnamefont{{Strolger}}},
  \bibinfo{author}{\bibfnamefont{J.}~\bibnamefont{{Tonry}}},
  \bibinfo{author}{\bibfnamefont{S.}~\bibnamefont{{Casertano}}},
  \bibinfo{author}{\bibfnamefont{H.~C.} \bibnamefont{{Ferguson}}},
  \bibinfo{author}{\bibfnamefont{B.}~\bibnamefont{{Mobasher}}},
  \bibinfo{author}{\bibfnamefont{P.}~\bibnamefont{{Challis}}},
  \bibinfo{author}{\bibfnamefont{A.~V.} \bibnamefont{{Filippenko}}},
  \bibinfo{author}{\bibfnamefont{S.}~\bibnamefont{{Jha}}},
  \bibinfo{author}{\bibfnamefont{W.}~\bibnamefont{{Li}}}, \bibnamefont{et~al.},
  \bibinfo{journal}{\apj} \textbf{\bibinfo{volume}{607}}, \bibinfo{pages}{665}
  (\bibinfo{year}{2004}), \eprint{arXiv:astro-ph/0402512}.

\bibitem[{\citenamefont{{Hoyle} and {Fowler}}(1960)}]{hoyle60}
\bibinfo{author}{\bibfnamefont{F.}~\bibnamefont{{Hoyle}}} \bibnamefont{and}
  \bibinfo{author}{\bibfnamefont{W.~A.} \bibnamefont{{Fowler}}},
  \bibinfo{journal}{\apj} \textbf{\bibinfo{volume}{132}}, \bibinfo{pages}{565}
  (\bibinfo{year}{1960}).

\bibitem[{\citenamefont{{Kramer}}(2010)}]{kramerprivate}
\bibinfo{author}{\bibfnamefont{M.}~\bibnamefont{{Kramer}}},
  \bibinfo{howpublished}{private communication} (\bibinfo{year}{2010}).

\bibitem[{\citenamefont{{Goriely}
  et~al.}(2011{\natexlab{a}})\citenamefont{{Goriely}, {Chamel}, {Janka}, and
  {Pearson}}}]{2011A&A...531A..78G}
\bibinfo{author}{\bibfnamefont{S.}~\bibnamefont{{Goriely}}},
  \bibinfo{author}{\bibfnamefont{N.}~\bibnamefont{{Chamel}}},
  \bibinfo{author}{\bibfnamefont{H.-T.} \bibnamefont{{Janka}}},
  \bibnamefont{and} \bibinfo{author}{\bibfnamefont{J.~M.}
  \bibnamefont{{Pearson}}}, \bibinfo{journal}{\aap}
  \textbf{\bibinfo{volume}{531}}, \bibinfo{pages}{A78+}
  (\bibinfo{year}{2011}{\natexlab{a}}), \eprint{1105.2453}.

\bibitem[{\citenamefont{{Goriely}
  et~al.}(2011{\natexlab{b}})\citenamefont{{Goriely}, {Bauswein}, and {-Thomas
  Janka}}}]{2011arXiv1107.0899G}
\bibinfo{author}{\bibfnamefont{S.}~\bibnamefont{{Goriely}}},
  \bibinfo{author}{\bibfnamefont{A.}~\bibnamefont{{Bauswein}}},
  \bibnamefont{and} \bibinfo{author}{\bibfnamefont{H.}~\bibnamefont{{-Thomas
  Janka}}}, \bibinfo{journal}{ArXiv e-prints}
  (\bibinfo{year}{2011}{\natexlab{b}}), \eprint{1107.0899}.

\bibitem[{\citenamefont{{Pearson} et~al.}(2011)\citenamefont{{Pearson},
  {Goriely}, and {Chamel}}}]{2011PhRvC..83f5810P}
\bibinfo{author}{\bibfnamefont{J.~M.} \bibnamefont{{Pearson}}},
  \bibinfo{author}{\bibfnamefont{S.}~\bibnamefont{{Goriely}}},
  \bibnamefont{and} \bibinfo{author}{\bibfnamefont{N.}~\bibnamefont{{Chamel}}},
  \bibinfo{journal}{\prc} \textbf{\bibinfo{volume}{83}},
  \bibinfo{pages}{065810} (\bibinfo{year}{2011}).

\bibitem[{\citenamefont{{Nauenberg}}(2008)}]{nauenberg08}
\bibinfo{author}{\bibfnamefont{M.}~\bibnamefont{{Nauenberg}}},
  \bibinfo{journal}{Journal for the History of Astronomy}
  \textbf{\bibinfo{volume}{39}}, \bibinfo{pages}{297} (\bibinfo{year}{2008}).

\bibitem[{\citenamefont{{Heilbron}}(1983)}]{heilbron83}
\bibinfo{author}{\bibfnamefont{J.~L.} \bibnamefont{{Heilbron}}},
  \bibinfo{journal}{Historical Studies in the Physical Sciences}
  \textbf{\bibinfo{volume}{13}}, \bibinfo{pages}{261} (\bibinfo{year}{1983}).

\end{thebibliography}

%%%%%%%%%%%%%%%%%%%%%%%%%%%%%%%%%%%%%%%%%%%%%%%%%%%%%%%%%%%%%%

\end{document}